\documentclass[a4paper]{article}

\usepackage[english]{babel}
\usepackage[utf8x]{inputenc}
\usepackage[T1]{fontenc}

\usepackage[a4paper,top=3cm,bottom=2cm,left=3cm,right=3cm,marginparwidth=1.75cm]{geometry}

\usepackage{amsmath}
\usepackage{authblk}
\usepackage{bm}
\usepackage{graphicx}
\usepackage{caption}  
\usepackage{booktabs}
\usepackage{algorithm}
\usepackage{algpseudocode}
\providecommand{\keywords}[1]{\textbf{\textit{Keywords:}} #1}

\title{Generalized simultaneous component analysis of binary and quantitative data}
\author{Yipeng Song\,$^{\text{1}}$, Johan A. Westerhuis\,$^{\text{1}}$, Nanne Aben\,$^{\text{2}}$, Lodewyk F.A. Wessels\,$^{\text{2}}$, Patrick J.F. Groenen\,$^{\text{3}}$ and Age K. Smilde\,$^{\text{}}$}

\affil[1]{Swammerdam Institute for Life Sciences, University of Amsterdam}
\affil[2]{Division of Molecular Carcinogenesis, Oncode Institute, Netherlands Cancer Institute}
\affil[3]{Econometric Institute, Erasmus School of Economics, Erasmus University}

\date{}

\begin{document}

\maketitle

\begin{abstract}
In the current era of systems biological research there is a need for the integrative analysis of binary and quantitative genomics data sets measured on the same objects. One standard tool of exploring the underlying dependence structure present in multiple quantitative data sets is simultaneous component analysis (SCA) model. However, it does not have any provisions when a part of the data are binary. To this end, we propose the generalized SCA (GSCA) model, which takes into account the distinct mathematical properties of binary and quantitative measurements in the maximum likelihood framework. Like in the SCA model, a common low dimensional subspace is assumed to represent the shared information between these two distinct types of measurements. However, the GSCA model can easily be overfitted when a rank larger than one is used, leading to some of the estimated parameters to become very large. To achieve a low rank solution and combat overfitting, we propose to use a concave variant of the nuclear norm penalty. An efficient majorization algorithm is developed to fit this model with different concave penalties. Realistic simulations (low signal-to-noise ratio and highly imbalanced binary data) are used to evaluate the performance of the proposed model in recovering the underlying structure. Also, a missing value based cross validation procedure is implemented for model selection. We illustrate the usefulness of the GSCA model for exploratory data analysis of quantitative gene expression and binary copy number aberration (CNA) measurements obtained from the GDSC1000 data sets.
\end{abstract}

\keywords{Data integration, SCA, binary data, low rank matrix approximation, concave penalty, majorization.}

\section{Introduction}
In biological research it becomes increasingly common to have measurements of different aspects of information on the same objects to study complex biological systems. The resulting coupled data sets should be analyzed simultaneously to explore the dependency between variables in different data sets and to reach a global understanding of the underlying biological system. The Simultaneous Component Analysis (SCA) model is one of the standard methods for the integrative analysis of such coupled data sets in different areas, from psychology to chemistry and biology \cite{van2009structured}. SCA discovers the common low dimensional column subspace of the coupled quantitative data sets, and this subspace represents the shared information between them.\\

Next to the quantitative measurements (such as gene expression data), it is common in biological research to have additional binary measurements, in which distinct categories differ in quality rather than in quantity (such as mutation data). Typical examples include the measurements of point mutations, which reflect the mutation status of the DNA sequence, the binary measurements of copy number aberrations (CNA), in which ``1'' indicates aberrations (gains or losses of segments in chromosomal regions) that occurred and ``0'' indicates the normal wild types status, and binarized DNA methylation measurements, in which ``1'' indicates a high level of methylation and ``0'' means a low level \cite{iorio2016landscape}. Compared to quantitative measurement, a binary measurement only has two mutually exclusive outcomes, such as presence vs absence (or true vs false), which are usually labeled as ``1'' and ``0''. However, ``1'' and ``0'' indicate abstract representations of two categories rather than quantitative values 1 and 0. As such, the special mathematical properties of binary data should be taken into account in the data analysis. In most biological data sets, the number of ``0''s is significantly larger than the number of ``1''s for most binary variables making the data imbalanced. Therefore, an additional requirement of the data analysis method is that it should be able to handle imbalanced data.\\

There is a need for statistical methods appropriate for doing an integrative analysis of coupled binary and quantitative data sets in biology research. The standard SCA models \cite{van2009structured, van2009integrating} that use column centering processing steps and least-squares loss criteria are not appropriate for binary data sets. Recently, iClusterPlus \cite{mo2013pattern} was proposed as a factor analysis framework to model discrete and quantitative data sets simultaneously by exploiting the properties of exponential family distributions. In this framework, the special properties of binary, categorical, and count variables are taken into account in a similar way as in generalized linear models. The common low dimensional latent variables and data set specific coefficients are used to fit the discrete and quantitative data sets. For the binary data set, the Bernoulli distribution is assumed and the canonical logit link function is used. The sum of the log likelihood is then used as the objective function. Furthermore, the approach allows the use of a lasso type penalty for feature selection. The Monte Carlo Newton–Raphson algorithm for this general framework, however, involves a very slow Markov Chain Monte Carlo simulation process. Both the high complexity of the model and the algorithmic inefficiency limit its use for large data sets and exploring its properties through simulations.\\

In this paper, we generalize the SCA model to binary and quantitative data from a probabilistic perspective similar as in Collins \cite{collins2002generalization} and Mo \cite{mo2013pattern}. However, the generalized SCA model can easily lead to overfitting by using a rank restriction higher larger than $1$, leading to some of the parameters to become very large. Therefore, a penalty on the singular values of the matrix contains parameters is used to simultaneously induce the low rank structure in a soft manner and to control the scale of estimated parameters. A natural choice is the convex nuclear norm penalty, which is widely used in low rank approximation problems \cite{koltchinskii2011nuclear, groenen2016multinomial, wu2015fast}. However, nuclear norm penalty shrinks all the singular values (latent factors) to the same degree, leading to biased estimates of the important latent factors. Hence, we would like to reduce the shrinkage for the most important latent factors while increase the shrinkage for unimportant latent factors. This nonlinear shrinkage strategy has shown its superiority in the recent work of low rank matrix approximation problems under the presence of Gaussian noise \cite{gavish2017optimal, josse2016adaptive}. Therefore, we will explore the nonlinear shrinkage of the latent factors through concave penalties in our GSCA model. The fitting of the resulting GSCA model is a penalized maximum likelihood estimation problem. We derive a Majorization-Minimization (MM) \cite{de1994block,hunter2004tutorial} based algorithm to solve it. Simple closed form updates for all the parameters are derived in each iteration. A missing value based cross validation procedure is also implemented to do model selection. Our algorithm is easy to implement and guaranteed to decrease the loss function monotonically in each iteration.\\

In the next sections, we will generalize the SCA model for binary and quantitative data, introduce the concave penalties and describe the majorization algorithm to estimate the model parameters. Section 4 introduces the simulations using low signal-to-noise ratios and highly imbalanced binary data, the performance of the GSCA model in recovering the underlying structure and the cross validation procedure. Section 5 introduces the GDSC data \cite{iorio2016landscape}, and the results of the analysis of this data.\\

\section{The GSCA model}
Before the GSCA model is introduced, consider the standard SCA model. The quantitative measurements on the same $I$ objects from two different platforms result into two data sets $\mathbf{X}_1$($I\times J_1$) and $\mathbf{X}_2$($I\times J_2$), in which $J_1$ and $J_2$ are the number of variables. Assume both $\mathbf{X}_1$ and $\mathbf{X}_2$ are column centered. The standard SCA model can be expressed as
\begin{equation}\label{eq1}
\begin{aligned}
\mathbf{X}_1 &= \mathbf{AB}_1^{\text{T}} + \mathbf{E}_1\\
\mathbf{X}_2 &= \mathbf{AB}_2^{\text{T}} + \mathbf{E}_2,\\
\end{aligned}
\end{equation}
where $\mathbf{A}$($I\times R$) denotes the common component scores (or latent variables), which span the common column subspace of $\mathbf{X}_1$ and $\mathbf{X}_2$, $\mathbf{B}_1$($J_1\times R$) and $\mathbf{B}_2$($J_2\times R$) are the data set specific loading matrices for $\mathbf{X}_1$ and $\mathbf{X}_2$ respectively, $\mathbf{E}_1$($I\times J_1$) and $\mathbf{E}_2$($I\times J_2$) are residuals, $R$, $R \ll \{I,J_1,J_2\}$, is an unknown low rank. Orthogonality is imposed on $\mathbf{A}$ as $\mathbf{A}^{\text{T}}\mathbf{A}=\mathbf{I}_{R}$, where $\mathbf{I}_{R}$ indicates the $R\times R$ identity matrix, to have a unique solution. $\mathbf{A}$, $\mathbf{B}_1$ and $\mathbf{B}_2$ are estimated by minimizing the sum of the squared residuals $\mathbf{E}_1$ and $\mathbf{E}_2$.\\

\subsection{The GSCA model of binary and quantitative data sets}
Following the probabilistic interpretation of the PCA model \cite{tipping1999probabilistic}, the high dimensional quantitative data set $\mathbf{X}_2$ can be assumed to be a noisy observation from a deterministic low dimensional structure $\mathbf{\Theta}_2$($I\times J_2$) with independent and identically distributed measurement noise, $\mathbf{X}_2 = \mathbf{\Theta}_2 + \mathbf{E}_2$. Elements in $\mathbf{E}_2$($I\times J_2$) follow a normal distribution with mean 0 and variance $\sigma^2$, $\epsilon_{2ij} \sim \text{N}(0,\sigma^2$). In the same way, following the interpretation of the exponential family PCA on binary data \cite{collins2002generalization}, we assume there is a deterministic low dimensional structure $\mathbf{\Theta}_1$($I\times J_1$) underlying the high dimensional binary observation $\mathbf{X}_1$. Elements in $\mathbf{X}_1$ follow the Bernoulli distribution with parameters $\phi(\mathbf{\Theta}_1)$, $x_{1ij} \sim \text{Ber}(\phi(\theta_{1ij}))$. Here $\phi()$ is the element wise inverse link function in the generalized linear model for binary data; $x_{1ij}$ and $\theta_{1ij}$ are the $ij$-th element of $\mathbf{X}_1$ and $\mathbf{\Theta}_1$ respectively. If the logit link is used, $\phi(\theta) = (1+\exp(-\theta))^{-1}$, while if the probit link is used, $\phi(\theta) = \Phi(\theta)$, where $\Phi$ is the cumulative density function of the standard normal distribution. Although in our paper, we only use the logit link in deriving the algorithm and in setting up the simulations, the option for the probit link is included in our implementation. The two link functions are similar, but their interpretations can be quite different \cite{agresti2013categorical}.\\

In the same way as in the standard SCA model, $\mathbf{\Theta}_1$ and $\mathbf{\Theta}_2$ are assumed to lie in the same low dimensional subspace, which represents the shared information between the coupled matrices $\mathbf{X}_1$ and $\mathbf{X}_2$. The commonly used column centering is not appropriate for the binary data set as the centered binary data will not be ``1'' and ``0'' anymore. Therefore, we include column offset terms $\bm{\mu}_1$($J_1 \times 1$) and $\bm{\mu}_2$($J_2 \times 1$) for a model based centering. The above ideas are modeled as
\begin{equation}\label{eq2}
\begin{aligned}
\mathbf{\Theta}_1 &= \mathbf{1}\bm{\mu}_1^{\text{T}} + \mathbf{AB}_1^{\text{T}}\\
\mathbf{\Theta}_2 &= \mathbf{1}\bm{\mu}_2^{\text{T}} + \mathbf{AB}_2^{\text{T}},\\
\end{aligned}
\end{equation}
where, $\mathbf{1}$($I\times 1$) is a $I$ dimensional vector of ones; the parameters $\mathbf{A}$, $\mathbf{B}_1$ and $\mathbf{B}_2$ have the same meaning as in the standard SCA model. Constraints $\mathbf{A}^{\text{T}}\mathbf{A}=I\mathbf{I}_R$ and $\mathbf{1}^{\text{T}}\mathbf{A} = \mathbf{0}$ are imposed to have a unique solution.\\

For the generalization to quantitative and binary coupled data, we follow the maximum likelihood estimation framework. The negative log likelihood for fitting coupled binary $\mathbf{X}_1$ and quantitative $\mathbf{X}_2$ is used as the objective function. In order to implement missing value based cross validation procedure \cite{bro2008cross}, we introduce two weighting matrices $\mathbf{Q}_1$($I\times J_1$) and $\mathbf{Q}_2$($I\times J_2$) to handle the missing elements. The $ij$-th element of $\mathbf{Q}_1$, $q_{1ij}$ equals 0 if the $ij$-th element in $\mathbf{X}_1$ is missing, while it equals 1 \textit{vice versa}. The same rules apply to $\mathbf{Q}_2$ and $\mathbf{X}_2$. The loss functions $f_1(\mathbf{\Theta}_1)$ for fitting $\mathbf{X}_1$ and $f_2(\mathbf{\Theta}_2,\sigma^2)$ for fitting $\mathbf{X}_2$ are defined as follows:
\begin{equation}\label{eq3}
\begin{aligned}
f_1(\mathbf{\Theta}_1) &= -\sum_{i}^{I}\sum_{j}^{J_1} q_{1ij} \left[x_{1ij}\log(\phi(\theta_{1ij})) + (1-x_{1ij})\log(1-\phi(\theta_{1ij}))\right] \\
f_2(\mathbf{\Theta}_2,\sigma^2)  &= \frac{1}{2\sigma^2}
              ||\mathbf{Q}_2 \odot (\mathbf{X}_2-\mathbf{\Theta}_2)||_F^2 +     \frac{1}{2} ||\mathbf{Q}_2||_0 \log(2\pi \sigma^2),\\
\end{aligned}
\end{equation}
where $\odot$ indicates element-wise multiplication; $||\quad||_F$ is the Frobenius norm of a matrix; $||\quad||_0$ is the pseudo $L_0$ norm of a matrix, which equals the number of nonzero elements.\\

The shared information between $\mathbf{X}_1$ and $\mathbf{X}_2$ is assumed to be fully represented by the low dimensional subspace spanned by the common component score matrix $\mathbf{A}$. Thus, $\mathbf{X}_1$ and $\mathbf{X}_2$ are conditionally independent given that the low dimensional structures $\mathbf{\Theta}_1$ and $\mathbf{\Theta}_2$ lie in the same low dimensional subspace. Therefore, the joint loss function is the direct sum of the negative log likelihood functions for fitting $\mathbf{X}_1$ and $\mathbf{X}_2$.
\begin{equation}\label{eq4}
\begin{aligned}
f(\mathbf{\Theta}_1,\mathbf{\Theta}_2,\sigma^2) &= -\log(p(\mathbf{X}_1,\mathbf{X}_2|\mathbf{\Theta}_1,\mathbf{\Theta}_2,\sigma^2))\\
& = -\log(p(\mathbf{X}_1|\mathbf{\Theta}_1) p(\mathbf{X}_2|\mathbf{\Theta}_2,\sigma^2))\\
              &= -\log(p(\mathbf{X}_1|\mathbf{\Theta}_1)-\log(p(\mathbf{X}_2|\mathbf{\Theta}_2,\sigma^2))\\
              &= f_1(\mathbf{\Theta}_1) + f_2(\mathbf{\Theta}_2,\sigma^2).\\
\end{aligned}
\end{equation}

\subsection{Concave penalties as surrogates for low rank constraint}
To arrive at meaningful solutions for the GSCA model, it is necessary to introduce penalties on the estimated parameters. If we take $\mathbf{\Theta} = [\mathbf{\Theta}_1 ~ \mathbf{\Theta}_2]$, $\bm{\mu} = [\bm{\mu}_1^{\text{T}} \bm{\mu}_2^{\text{T}}]^{\text{T}}$, and $\mathbf{B} = [\mathbf{B}_1 ~ \mathbf{B}_2]$, equation (\ref{eq2}) in the GSCA model can be expressed as $\mathbf{\Theta} = \mathbf{1}\bm{\mu}^{\text{T}} + \mathbf{AB}^{\text{T}}$. In the above interpretation of the GSCA model, the low rank constraint on the column centered $\mathbf{\Theta}$ is expressed as the multiplication of two rank $R$ matrices $\mathbf{A}$, $\mathbf{B}$, $\mathbf{Z} = \mathbf{\Theta} - \mathbf{1}\bm{\mu}^{\text{T}} = \mathbf{AB}^{\text{T}}$. However, using a exact low rank constraint in the GSCA model has some issues. First, the maximum likelihood estimation of this model easily leads to overfitting. Given the constraint that $\mathbf{A}^{\text{T}}\mathbf{A}=I\mathbf{I}$, overfitting represents itself in a way that some elements in $\mathbf{B}_1$ tend to diverge to plus or minus infinity. In addition, the exact low rank $R$ in the GSCA model is commonly unknown and its selection is not straightforward.\\

In this paper, we take a penalty based approach to control the scale of estimated parameters and to induce a low rank structure simultaneously. The low rank constraint on $\mathbf{Z}$ is obtained by a penalty function $g(\mathbf{Z})$, which shrinks the singular values of $\mathbf{Z}$ to achieve a low rank structure. The most widely used convex surrogate of a low rank constraint is the nuclear norm penalty, which is simply the sum of singular values, $g(\mathbf{Z}) = \sum_{r} \xi_r(\mathbf{Z})$ \cite{koltchinskii2011nuclear}, where $\xi_r(\mathbf{Z})$ represents the $r$-th singular value of $\mathbf{Z}$. The nuclear norm penalty was also used in a related work \cite{wu2015fast}. Although the convex nuclear norm penalty is easy to optimize, the same amount of shrinkage is applied to all the singular values, leading to biased estimates of the large singular values. Recent work \cite{gavish2017optimal, lu2015generalized} already showed the superiority of concave surrogates of a low rank constraint under Gaussian noise compared to the nuclear norm penalty. We take $g(\mathbf{Z}) = \sum_{r} g(\xi_r(\mathbf{Z}))$ as our concave surrogate of a low rank constraint on $\mathbf{Z}$, where $g(\xi_r)$ is a concave penalty function of $\xi_r$. After replacing the low rank constraint in equation (\ref{eq4}) by $g(\mathbf{Z})$, our model becomes,
\begin{equation}\label{eq5}
\begin{aligned}
    \min_{\bm{\mu},\mathbf{Z},\sigma^2} \quad & f_1(\mathbf{\Theta}_1) + f_2(\mathbf{\Theta}_2,\sigma^2) + \lambda g(\mathbf{Z}) \\
    \text{s.t.~} \mathbf{\Theta} &= \mathbf{1}\bm{\mu}^{\text{T}} + \mathbf{Z} \\
     \mathbf{\Theta} &= [\mathbf{\Theta}_1 ~ \mathbf{\Theta}_2] \\
     \mathbf{1}^{\text{T}}\mathbf{Z} &= \mathbf{0}.
\end{aligned}
\end{equation}

The most commonly used non-convex surrogates of a low rank constraint are concave functions, including $L_{q:0 < q < 1}$ (bridge penalty) \cite{fu1998penalized,liu2007support}, smoothly clipped absolute deviation (SCAD) \cite{fan2001variable}, a frequentist version of the generalized double Pareto (GDP) shrinkage \cite{armagan2013generalized} and others \cite{lu2015generalized}. We include the first three concave penalties in our algorithm. Their formulas and supergradients (the counterpart concept of subgradient in convex analysis, which will be used in the derivation of the algorithm) are shown in Table \ref{tab1}, and their thresholding properties are shown in Fig.~1. Since the nuclear norm penalty is a linear function of the singular values, it is both convex and concave. Also, it is a special case of $L_q$ penalty when setting $q=1$. Thus, the algorithm developed in this paper also applies to nuclear norm penalty.\\

\begin{table}[htbp]
\centering
\caption{\label{tab1} Some commonly used concave penalty functions. $\eta$ is taken as the singular value and $q$, $\lambda$ and $\gamma$ are tuning parameters.}
\begin{tabular}{lll}
  \toprule
Penalty & Formula & Supergradient \\
  \midrule
 Nuclear norm & $ \lambda \eta $ & $\lambda$ \\

$L_{q}$ & $ \lambda \eta^q $ & $\left\{ \begin{array}{ll} +\infty &\textrm{$\eta=0$}\\
                                 \lambda q \eta^{q-1} &\textrm{$\eta>0$}\\ \end{array} \right.$ \\

SCAD & $\left\{ \begin{array}{ll} \lambda \eta &\textrm{$\eta \leq \lambda$}\\
 \frac{-\eta^2+2\gamma \lambda \eta - \lambda^2}{2(\gamma-1)} &\textrm{$\lambda < \eta \leq \gamma \lambda$}\\
 \frac{\lambda^2(\gamma+1)}{2} &\textrm{$\eta > \gamma \lambda$}\\ \end{array} \right.$ &
                          $\left\{ \begin{array}{ll} \lambda &\textrm{$\eta \leq \lambda$}\\
 \frac{\gamma \lambda - \eta}{\gamma-1} &\textrm{$\lambda < \eta \leq \gamma \lambda$}\\
 0 &\textrm{$\eta > \gamma \lambda$}\\ \end{array} \right.$ \\

GDP & $ \lambda \log(1+\frac{\eta}{\gamma}) $ & $\frac{\lambda}{\gamma + \eta}$ \\
  \bottomrule
\end{tabular}
\end{table}

\begin{figure}[h!]\label{Fig:1}
    \centering
    \includegraphics[width=\textwidth]{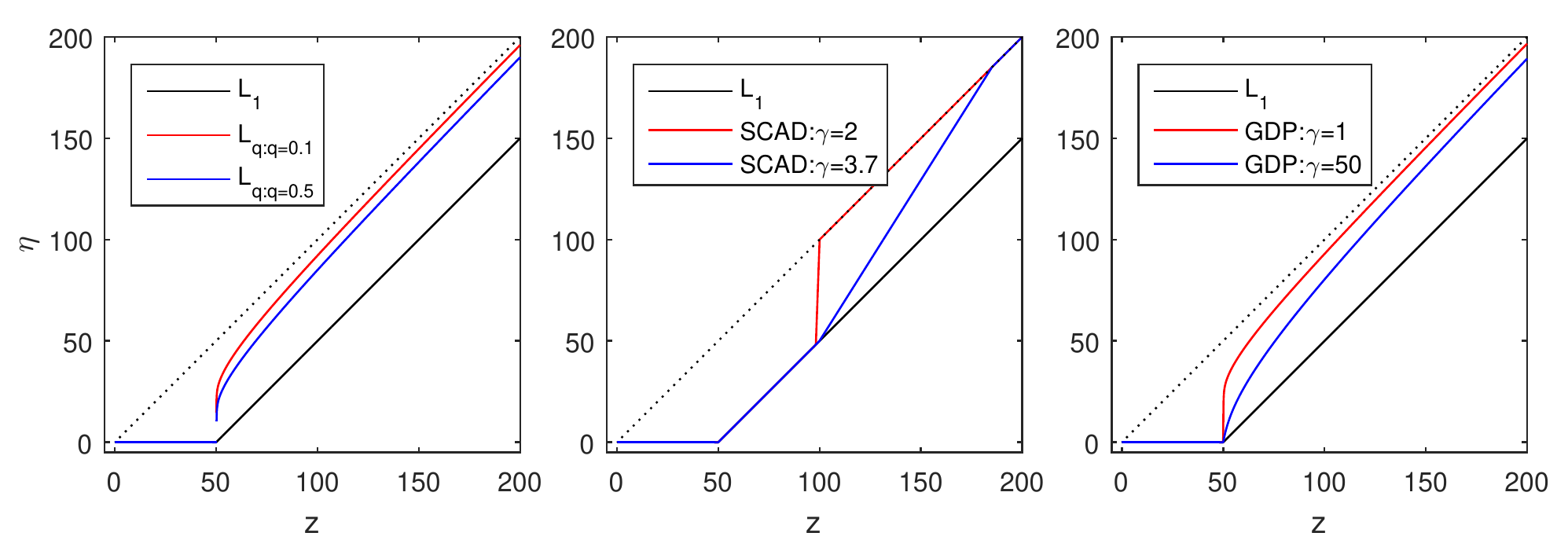}
    \caption*{\textbf{Fig.~1} Thresholding properties of the $L_{q}$, SCAD and GDP penalties when the same degree of shrinkage is achieved. $L_{1}$: nuclear norm penalty. $z$ indicates the original singular values while $\eta$ is the value after thresholding. Not that in contrast to SCAD and GDP, the $L_{q: 0 < q < 1}$ penalty has a small discontinuity region, thus continuous thresholding can not be achieved.}
\end{figure}

\section{Algorithm}
Parameters $\bm{\mu}$, $\mathbf{Z}$, and $\sigma^2$ of the joint loss function in equation (\ref{eq5}) are updated alternatingly while fixing other parameters until reaching predefined stopping criteria. If the updating sequence follows the same order, the joint loss function is guaranteed to decrease monotonically. However, even when fixing parameters $\bm{\mu}$ and $\sigma^2$, the minimization of equation (\ref{eq5}) with respect to $\mathbf{Z}$ is still a non-smooth and non-convex problem. We solve this problem using the MM principle \cite{de1994block,hunter2004tutorial}.\\

\subsection{The majorization of $f_1(\mathbf{\Theta}_1) + f_2(\mathbf{\Theta}_2,\sigma^2) + \lambda g(\mathbf{Z})$}
When fixing $\sigma^2$, we can majorize $f(\mathbf{\Theta}) = f_1(\mathbf{\Theta}_1) + f_2(\mathbf{\Theta}_2)$ to a quadratic function of the parameter $\mathbf{\Theta}$. In addition, the concave penalty function $g(\mathbf{Z})$ can be majorized to a linear function of the singular values by exploiting the concavity. The resulting majorized problem can be analytically solved by weighted singular value thresholding \cite{lu2015generalized}. In the following derivation, the symbol $c$ represents a constant doesn't depend on any unknown parameters, rather than a specific value.\\

\subsubsection*{The majorization of $f(\mathbf{\Theta})$}
Both $f_1(\mathbf{\Theta}_1)$ and $f_2(\mathbf{\Theta}_2)$ can be expressed as $f_1(\mathbf{\Theta}_1) = \sum_{i}^{I}\sum_{j}^{J_1} q_{1ij} f_{1ij}(\theta_{1ij})$ and $f_2(\mathbf{\Theta}_2) = \sum_{i}^{I}\sum_{j}^{J_2} q_{2ij} f_{2ij}(\theta_{2ij})$, in which $f_{1ij}(\theta_{1ij}) = -\left[x_{1ij}\log(\phi(\theta_{1ij})) + (1-x_{1ij})\log(1-\phi(\theta_{1ij}))\right]$ and $f_{2ij}(\theta_{2ij}) = \frac{1}{2\sigma^2} (x_{2ij}-\theta_{2ij})^2 + c$. When logit link is used, the following results can be easily derived out, $\nabla f_{1ij}(\theta_{1ij}) = \phi(\theta_{1ij}) - x_{1ij}$, $\nabla^2 f_{1ij}(\theta_{1ij}) = \phi(\theta_{1ij})(1-\phi(\theta_{1ij}))$, $\nabla f_{2ij}(\theta_{2ij}) = \frac{1}{\sigma^2} (\theta_{2ij} - x_{2ij})$, $\nabla^2 f_{2ij}(\theta_{2ij}) = \frac{1}{\sigma^2}$. Assume that both $\nabla^2 f_{1ij}(\theta_{1ij})$ and $\nabla^2 f_{2ij}(\theta_{2ij})$ are upper bounded by a constant $L$. Since $\nabla^2 f_{1ij}(\theta_{1ij}) \leq 0.25$ when logit link is used \cite{de2006principal}, we can set $L=\text{max}(0.25, 1/\sigma^2)$. Take $f(\theta)$ as the general representation of $f_{1ij}(\theta_{1ij})$ and $f_{2ij}(\theta_{2ij})$. According to the Taylor's theorem and the assumption that $\nabla^2 f(\theta) \leq L$ for $\theta \in \text{domain}f$, we have the following inequality,
\begin{equation}\label{eq6}
\begin{aligned}
f(\theta) &= f(\theta^k) + <\nabla f(\theta^k), \theta-\theta^k> + \frac{1}{2}(\theta-\theta^k)^{\text{T}} \nabla^{2}f(\theta^k + t(\theta-\theta^k))(\theta-\theta^k) \\
          &\leq f(\theta^k) + <\nabla f(\theta^k), \theta-\theta^k> + \frac{L}{2}(\theta-\theta^k)^2 \\
          &= \frac{L}{2}(\theta-\theta^k + \frac{1}{L}\nabla f(\theta^k))^2 + c,\\
\end{aligned}
\end{equation}
where $\theta^k$ is the $k$-th approximation of $\theta$, $t$ is an unknown constant and $t\in[0,1]$. Therefore, we have the following inequalities about $f_{1ij}(\theta_{1ij})$ and $f_{2ij}(\theta_{2ij})$,
$f_{1ij}(\theta_{1ij}) \leq \frac{L}{2}(\theta_{1ij} - \theta_{1ij}^k + \frac{1}{L}\nabla f_{1ij}(\theta_{1ij}^k ))^2 + c$ and $f_{2ij}(\theta_{2ij}) \leq \frac{L}{2}(\theta_{2ij} - \theta_{2ij}^k + \frac{1}{L} \nabla f_{2ij}(\theta_{2ij}^k))^2 + c$.\\

Assume $\nabla f_1(\mathbf{\Theta}_1^k)$ and $\nabla f_1(\mathbf{\Theta}_1^k)$ are the matrix forms of $\nabla f_{1ij}(\theta_{1ij}^k )$ and $\nabla f_{2ij}(\theta_{2ij}^k )$ respectively. The inequality of $f_1(\mathbf{\Theta}_1)$ can be derived out as $f_1(\mathbf{\Theta}_1)\leq \frac{L}{2} \sum_{i}^{I}\sum_{j}^{J_1} q_{1ij}[(\theta_{1ij} - \theta_{1ij}^k + \frac{1}{L}\nabla f_{1ij}(\theta_{1ij}^k ))^2] + c = \frac{L}{2} ||\mathbf{Q}_1 \odot (\mathbf{\Theta}_1 - \mathbf{\Theta}_1^k + \frac{1}{L} \nabla f_1(\mathbf{\Theta}_1^k))||_F^2 + c$. In the same way, the inequality of $f_2(\mathbf{\Theta}_2)$ is $f_2(\mathbf{\Theta}_2)\leq \frac{L}{2} ||\mathbf{Q}_2 \odot (\mathbf{\Theta}_2 - \mathbf{\Theta}_2^k + \frac{1}{L} \nabla f_2(\mathbf{\Theta}_2^k))||_F^2 + c$. Based on these two inequalities, we can derive out the upper bound of $f(\mathbf{\Theta})$ at the $k$-th approximated parameter $\mathbf{\Theta}^k = [\mathbf{\Theta}_1^k ~ \mathbf{\Theta}_2^k]$ as follows,
\begin{equation}\label{eq7}
\begin{aligned}
f(\mathbf{\Theta}) &= f_1(\mathbf{\Theta}_1) + f_2(\mathbf{\Theta}_2)\\
                   &\leq \frac{L}{2} ||\mathbf{Q}_1 \odot (\mathbf{\Theta}_1 - \mathbf{\Theta}_1^k + \frac{1}{L} \nabla f_1(\mathbf{\Theta}_1^k))||_F^2 + \frac{L}{2} ||\mathbf{Q}_2 \odot (\mathbf{\Theta}_2 - \mathbf{\Theta}_2^k + \frac{1}{L} \nabla f_2(\mathbf{\Theta}_2^k))||_F^2 + c\\
                   & = \frac{L}{2} ||\mathbf{Q} \odot(\mathbf{\Theta} - \mathbf{\Theta}^k + \frac{1}{L} \nabla f(\mathbf{\Theta}^k))||_F^2 + c,
\end{aligned}
\end{equation}
where $\nabla f(\mathbf{\Theta}^k) = [\nabla f_1(\mathbf{\Theta}_1^k) ~ \nabla f_2(\mathbf{\Theta}_2^k)]$, $\nabla f_1(\mathbf{\Theta}_1^k) =  \phi(\mathbf{\Theta}_1^k - \mathbf{X}_1)$ and $\nabla f_2(\mathbf{\Theta}_2^k) = \frac{1}{\sigma^2} (\mathbf{\Theta}_2^k - \mathbf{X}_2)$. Following \cite{kiers1997weighted}, we further majorize the weighted least-squares in equation (\ref{eq7}) into a quadratic function of $\mathbf{\Theta}$ as
\begin{equation}\label{eq8}
\begin{aligned}
              &\frac{L}{2} ||\mathbf{Q} \odot(\mathbf{\Theta} - \mathbf{\Theta}^k + \frac{1}{L} \nabla f(\mathbf{\Theta}^k))||_F^2 \\
             &\leq  \frac{L}{2} ||\mathbf{\Theta}-\mathbf{H}^k||_F^2 + c, \\
\end{aligned}
\end{equation}
where $\mathbf{H}^k =  \mathbf{Q} \odot(\mathbf{\Theta}^k - \frac{1}{L} \nabla f(\mathbf{\Theta}^k)) + (\mathbf{1}\mathbf{1}^{\text{T}}-\mathbf{Q})\odot \mathbf{\Theta}^k = \mathbf{\Theta}^k - \frac{1}{L}(\mathbf{Q} \odot \nabla f(\mathbf{\Theta}^k))$.\\

\subsubsection*{The majorization of $g(\mathbf{Z})$}
Let $g(\xi_r)$ be a concave function. From the definition of concavity \cite{boyd2004convex}, we have $g(\xi_r) \leq g(\xi_r^k) + \omega_r^k(\xi_r - \xi_r^k) = \omega_r^k \xi_r + c$, in which $\xi_r^k = \xi_r(\mathbf{Z}^k)$ is the $r$-th singular value of the $k$-th approximation $\mathbf{Z}^k$ and $c$ is a constant doesn't depend on any unknown parameter. Also, $\omega_r^k \in \partial g(\xi_r^k)$ and $\partial g(\xi_r^k)$ is the set of supergradients of function $g()$ at $\xi_r^k$. For all the concave penalties used in our paper, their supergradient is unique, thus $\omega_r^k = \partial g(\xi_r^k)$. Therefore, $g(\mathbf{Z})= \sum_{r}g(\xi_r(Z))$ can be majorized as follows
\begin{equation}\label{eq9}
\begin{aligned}
g(\mathbf{Z}) &= \sum_{r}g(\xi_r(\mathbf{Z}))\\
              &\leq \sum_{r}\omega_{r}^k \xi_r(\mathbf{Z}) + c\\
       \omega_r^k &= \partial g(\xi_r(\mathbf{Z}^k)).
\end{aligned}
\end{equation}

\subsubsection*{The majorization of $f(\mathbf{\Theta}) + \lambda g(\mathbf{Z})$}
To summarize the above results, $f(\mathbf{\Theta}) + \lambda g(\mathbf{Z})$ has been majorized to the following function.
\begin{equation}\label{eq10}
\begin{aligned}
              & \frac{L}{2} ||\mathbf{\Theta} - \mathbf{H}^k||_F^2 + \lambda \sum_{r}\omega_{r}^k \xi_r(\mathbf{Z}) + c\\
        \mathbf{\Theta} &= \mathbf{1}\bm{\mu}^{\text{T}} + \mathbf{Z} \\
         \mathbf{H}^k &= \mathbf{\Theta}^k - \frac{1}{L}(\mathbf{Q} \odot \nabla f(\mathbf{\Theta}^k)) \\
              \omega_r^k &= \partial g(\xi_r(\mathbf{Z}^k))\\
              \mathbf{1}^{\text{T}}\mathbf{Z} &= \mathbf{0}.
\end{aligned}
\end{equation}

\subsection{Block coordinate descent}
We optimize $\bm{\mu}$, $\mathbf{Z}$ and $\sigma^2$ alternatingly while fixing the other parameters. However, updating $\bm{\mu}$ and $\mathbf{Z}$ depend on solving the majorized problem in equation (\ref{eq10}) rather than solving the original problem in equation (\ref{eq5}). Because of the MM principle, this step will also monotonically decrease the original loss function in equation (\ref{eq5}).\\

\subsubsection*{Updating $\bm{\mu}$}
The analytical solution of $\bm{\mu}$ in equation (\ref{eq10}) is simply the column mean of $\mathbf{H}^k$, $\bm{\mu} = \frac{1}{I} (\mathbf{H}^k)^{\text{T}} \mathbf{1}$.\\

\subsubsection*{Updating $\mathbf{Z}$}
After deflating the offset term $\bm{\mu}$, the loss function in equation (\ref{eq10}) becomes $\frac{L}{2} ||\mathbf{Z} - \mathbf{J} \mathbf{H}^k||_F^2 + \lambda \sum_{r}\omega_{r}^k \xi_r$, in which $\mathbf{J} = \mathbf{I} - \frac{1}{I} \mathbf{1} \mathbf{1}^{\text{T}}$ is the column centering matrix. The solution of the resulting problem is equivalent to the proximal operator of the weighted sum of singular values, which has an analytical form solution \cite{lu2015generalized}. Suppose $\mathbf{USV}^{\text{T}} = \mathbf{J} \mathbf{H}^k$ is the SVD decomposition of $\mathbf{J} \mathbf{H}^k$, the analytical form solution of $\mathbf{Z}$ is $\mathbf{Z} = \mathbf{US}_{\omega \lambda /L}\mathbf{V}^{\text{T}}$, in which $\mathbf{S}_{\omega \lambda /L} = \text{Diag}\{(s_{rr}-\lambda \omega_r /L)_{+}\}$ and $s_{rr}$ is the $i$-th diagonal element in $\mathbf{S}$.\\

\subsubsection*{Updating $\sigma^2$}
By setting the gradient of $f(\mathbf{\Theta},\sigma^2)$ in equation (\ref{eq5}) with respect to $\sigma^2$ to be 0, we have the following analytical solution of $\sigma^2$,  $\sigma^2= \frac{1}{||\mathbf{Q}_2||_0} ||\mathbf{Q}_2 \odot (\mathbf{X}_2 - \mathbf{\Theta}_2)||_F^2$. When no low rank estimation of $\mathbf{Z}$ can be achieved, the constructed model is close to a saturated model and the estimated $\hat{\sigma}^2$ is close to 0. In that case, when $\hat{\sigma}^2<0.05$, the algorithm stops and gives a warning that a low rank estimation has not been achieved.\\

\subsubsection*{Initialization and stopping criteria}
Random initialization is used. All the elements in $\mathbf{Z}^0$ are sampled from the standard uniform distribution, $\bm{\mu}^0$ is set to 0 and $(\sigma^2)^0$ is set to 1. The relative change of the objective value is used as the stopping criteria. Pseudocode of the algorithm described above is shown in Algorithm \ref{alg1}. $\epsilon_f$ is the tolerance of relative change of the loss function.\\

\begin{algorithm}[htb]
  \caption{A MM algorithm for fitting the GSCA model with concave penalties.}
  \label{alg1}
  \begin{algorithmic}[1]
    \Require
      $\mathbf{X}_1$, $\mathbf{X}_2$, penalty, $\lambda$, $\gamma$;
    \Ensure
      $\hat{\bm{\mu}}$, $\hat{\mathbf{Z}}$, $\hat{\sigma}^2$;
    \State Compute $\mathbf{Q}_1$, $\mathbf{Q}_2$ for missing values in $\mathbf{X}_1$ and $\mathbf{X}_2$, and $\mathbf{Q} = [\mathbf{Q}_1 ~ \mathbf{Q}_2]$;
    \State Initialize $\bm{\mu}^0$, $\mathbf{Z}^0$, $(\sigma^2)^0$;
    \State $k = 0$;
    \While{$(f^{k-1}-f^{k})/f^{k-1}>\epsilon_f$}
        \State $\nabla f_1(\mathbf{\Theta}_1^k) = \phi(\mathbf{\Theta}_1^k) - \mathbf{X}_1$; $\nabla f_2(\mathbf{\Theta}_2^k) = \frac{1}{(\sigma^2)^k} (\mathbf{\Theta}_2^k - \mathbf{X}_2)$;
        \State $\nabla f(\mathbf{\Theta}^k) = [\nabla f_1(\mathbf{\Theta}_1^k) ~ \nabla f_2(\mathbf{\Theta}_2^k)]$;
        \State $L_k=\text{max}(0.25,1/(\sigma^2)^k)$;
        \State $\mathbf{H}^{k} = \mathbf{\Theta}^{k}- \frac{1}{L_{k}} (\mathbf{Q} \odot \nabla f(\mathbf{\Theta}^{k}))$;
        \State $\omega_r^k = \partial g(\xi_r(\mathbf{Z}^k))$;
        \State $\bm{\mu}^{k+1} = \frac{1}{I} (\mathbf{H}^{k})^{\text{T}} \mathbf{1}$;
        \State $\mathbf{USV}^{\text{T}} = \mathbf{J}\mathbf{H}^{k}$;
        \State $\mathbf{S}_{\lambda \omega /L_{k}} = \text{Diag}\{ (s_{rr} - \lambda \omega_r^k /L_{k})_{+}\}$;
        \State $\mathbf{Z}^{k+1} = \mathbf{US}_{\lambda \omega/L_k}\mathbf{V}^{\text{T}}$;
        \State $\mathbf{\Theta}^{k+1} = \mathbf{1}(\bm{\mu}^{k+1})^{\text{T}} + \mathbf{Z}^{k+1}$;
        \State $[\mathbf{\Theta}_1^{k+1} ~ \mathbf{\Theta}_2^{k+1}] = \mathbf{\Theta}^{k+1}$;
        \State $(\sigma^2)^{k+1} = \frac{1}{||\mathbf{Q}_2||_0} ||\mathbf{Q}_2 \odot (\mathbf{X}_2 - \mathbf{\Theta}_2^{k+1})||_F^2$
        \State $k=k+1$;
    \EndWhile
  \end{algorithmic}
\end{algorithm}

\section{Simulation}
To see how well the GSCA model is able to reconstruct data generated according to the model, we do a simulation study with similar characteristics as a typical empirical data set. We first simulate the imbalanced binary $\mathbf{X}_1$ and quantitative $\mathbf{X}_2$ following the GSCA model with logit link and low signal-to-noise ratio (SNR). After that, we evaluate the GSCA model with respect to 1) the quality of the reconstructed low rank structure from the model, and 2) the reconstruction of true number of dimensions.\\

\subsection{Data generating process}
Motivated by \cite{davenport20141}, we define the SNR for generating binary data according to the latent variable interpretation of the generalized linear models of binary data. Elements in $\mathbf{X}_1$ are independent and indirect binary observations of the corresponding elements in an underlying quantitative matrix $\mathbf{X}_1^{\ast}$($I\times J_1$), $x_{1ij} = 1$ if $x_{1ij}^{\ast}>0$ and $x_{1ij} = 0$ otherwise. $\mathbf{X}_1^{\ast}$ can be expressed as $\mathbf{X}_1^{\ast} = \mathbf{\Theta}_1 + \mathbf{E}_1$, in which $\mathbf{\Theta}_1 = \mathbf{1}\bm{\mu}_1^{\text{T}} + \mathbf{AB}_1^{\text{T}}$, and elements in $\mathbf{E}_1$ follow the standard logistic distribution, $\epsilon_{1ij} \sim \text{Logistic}(0,1)$.
The SNR for generating binary data $\mathbf{X}_1$ is defined as $\text{SNR}_1 = ||\mathbf{AB}_1^{\text{T}}||_{F}^2/||\mathbf{E}_1||_{F}^2$. Assume the quantitative $\mathbf{X}_2$ is simulated as $\mathbf{X}_2 = \mathbf{\Theta}_2 + \mathbf{E}_2$, in which $\mathbf{\Theta}_2 = \mathbf{1}\bm{\mu}_2^{\text{T}} + \mathbf{AB}_2^{\text{T}}$ and elements in $\mathbf{E}_2$ follow a normal distribution with 0 mean and $\sigma^2$ variance, $\epsilon_{2ij} \sim N(0,\sigma^2)$. The SNR for generating quantitative $\mathbf{X}_2$ is defined as $\text{SNR}_2 = ||\mathbf{AB}_2^{\text{T}}||_{F}^2/||\mathbf{E}_2||_{F}^2$.\\

After the definition of the SNR, we simulate the coupled binary $\mathbf{X}_1$ and quantitative $\mathbf{X}_2$ as follows. $\bm{\mu}_1$ represents the logit transform of the marginal probabilities of binary variables and $\bm{\mu}_2$ represents the mean of the marginal distributions of quantitative variables. They will be simulated according to the characteristic of real biological data set. The score matrix $\mathbf{A}$ and loading matrices $\mathbf{B}_1$, $\mathbf{B}_2$ are simulated as follows. First, we express $\mathbf{A}\mathbf{B}_1^{\text{T}}$ and $\mathbf{A}\mathbf{B}_2^{\text{T}}$ in a SVD type as $\mathbf{A}\mathbf{B}_1^{\text{T}} = \mathbf{U}\mathbf{D}_1\mathbf{V}_1^{\text{T}}$ and $\mathbf{A}\mathbf{B}_2^{\text{T}} = \mathbf{U}\mathbf{D}_2\mathbf{V}_1^{\text{T}}$, in which $\mathbf{U}^{\text{T}}\mathbf{U} = \mathbf{I}_R$, $\mathbf{D}_1$ and $\mathbf{D}_2$ are diagonal matrices, $\mathbf{V}_1^{\text{T}}\mathbf{V}_1 = \mathbf{I}_R$ and $\mathbf{V}_2^{\text{T}}\mathbf{V}_2 = \mathbf{I}_R$. All the elements in $\mathbf{U}$, $\mathbf{V}_1$ and $\mathbf{V}_2$ are independently sampled from the standard normal distribution. Then, $\mathbf{U}$, $\mathbf{V}_1$ and $\mathbf{V}_2$ are orthogonalized by the QR algorithm. The diagonal matrix $\mathbf{D}$($R\times R$) is simulated as follows. $R$ elements are sampled from standard normal distribution, their absolute values are sorted in decreasing order. To satisfy the pre-specified $\text{SNR}_1$ and $\text{SNR}_2$, $\mathbf{D}$ is scaled by positive scalars $c_1$ and $c_2$ as $\mathbf{D}_1 = c_1\mathbf{D}$ and $\mathbf{D}_2 = c_2\mathbf{D}$. Then, binary elements in $\mathbf{X}_1$ are sampled from the Bernoulli distribution with corresponding parameter $\phi(\theta_{1ij})$, in which $\phi()$ is inverse logit function and $\mathbf{\Theta}_1 = \mathbf{1}\bm{\mu}_1^{\text{T}} + \mathbf{AB}_1^{\text{T}}$. Quantitative data set $\mathbf{X}_2$ is generated as $\mathbf{X}_2 = \mathbf{\Theta}_2 + \mathbf{E}_2$, in which $\mathbf{\Theta}_2 = \mathbf{1}\bm{\mu}_2^{\text{T}} + \mathbf{AB}_2^{\text{T}}$ and elements in $\mathbf{E}_2$ are sampled from $N(0,\sigma^2)$. Take $\mathbf{Z} = \mathbf{A}\mathbf{B}^{\text{T}}$, $\mathbf{B} = [\mathbf{B}_1 ~ \mathbf{B}_2]$. In order to make $\mathbf{1}^{\text{T}}\mathbf{Z} = \mathbf{0}$, we further deflate the column offset of $\mathbf{Z}$ to the simulated $\bm{\mu}$, $\bm{\mu} = [\bm{\mu}_1^{\text{T}} ~ \bm{\mu}_2^{\text{T}}]^{\text{T}}$. This step will not change the value of $\mathbf{\Theta}_1$ and $\mathbf{\Theta}_2$, thus doesn't effect the simulation of $\mathbf{X}_1$ and $\mathbf{X}_2$.\\

\subsection{Evaluation metric and model selection}
As for simulated data sets, the true parameters $\mathbf{\Theta} = [\mathbf{\Theta}_1~\mathbf{\Theta}_2]$, $\bm{\mu} = [\bm{\mu}_1^{\text{T}} \bm{\mu}_1^{\text{T}}]^{\text{T}}$ and $\mathbf{Z} = \mathbf{A}\mathbf{B}^{\text{T}}$ are available. Therefore, the generalization error of the constructed model can be evaluated by comparing the true parameters and their model estimates. Thus, the evaluation metric is defined as the relative mean squared error (RMSE) of the model parameters. The RMSE of estimating $\mathbf{\Theta}$ is defined as $\text{RMSE}(\mathbf{\Theta}) = ||\mathbf{\Theta}-\hat{\mathbf{\Theta}}||_F^2/||\mathbf{\Theta}||_F^2$, where $\mathbf{\Theta}$ represents the true parameter and $\hat{\mathbf{\Theta}}$ its GSCA model estimate. The RMSE of $\bm{\mu}$ and $\mathbf{Z}$, are expressed as $\text{RMSE}(\bm{\mu})$ and $\text{RMSE}(\mathbf{Z})$ and they are defined in the same way as for $\mathbf{\Theta}$.\\

For real data sets, K-fold missing value based cross validation (CV) is used to estimate the generalization error of the constructed model. To make the prediction of the left out fold elements independent to the constructed model based on the reminding folds, the data is partitioned into K folds of elements which are selected in a diagonal style rather than row wise from $\mathbf{X}_1$ and $\mathbf{X}_2$ respectively, similar to the leave out patterns described by Wold \cite{wold1978cross, bro2008cross}. The test set elements of each fold in $\mathbf{X}_1$ and $\mathbf{X}_2$ are taken as missing values, and the remaining data are used to construct a GSCA model. After estimation of $\hat{\mathbf{\Theta}}$ and $\hat{\sigma^2}$ are obtained from the constructed GSCA model, the negative log likelihood of using $\hat{\mathbf{\Theta}}$, $\hat{\sigma^2}$ to predict the missing elements (left out fold) is recorded. This negative log likelihood is scaled by the number of missing elements. This process is repeated K times until all the K folds have been left out once. The mean of the K scaled negative log likelihoods is taken as the CV error.\\

When we define $\mathbf{X}=[\mathbf{X}_1 ~ \mathbf{X}_2]$ and $J=J_1+J_2$, the penalty term $\lambda g(\mathbf{Z})$ is not invariant to the number of non-missing elements in $\mathbf{X}$, as the joint loss function (equation (\ref{eq4})) is the sum of the log likelihoods for fitting all the non-missing elements in the data $\mathbf{X}$. Therefore, we effectively follow a similar approach as Fan \cite{fan2001variable} by adjusting the penalty strength parameter $\lambda$ for the relative number observations. By setting one fold of elements to be missing during the CV process, $\lambda||\mathbf{X}||_0/(I\times J)$ rather than $\lambda$ is used as the amount of penalty. During the K-fold CV process, a warm start strategy, using the results of previous constructed model as the initialization of next model, is applied. In this way, the K-fold CV can be greatly accelerated. The speed of the GSCA models with different penalties using different stopping criteria, and the corresponding CV procedure, are fully characterized in Table S1. All the computations are performed on a laptop with an i5-5300U CPU, 8GB RAM, 64-bit Windows 10 system and MATLAB of R2015a.\\

In the model selection process, the tuning parameter $\lambda$ and hyper-parameters ($q$ in $L_{q}$ and $\gamma$ in SCAD and GDP) can be selected by a grid search. However, previous work of using these penalty functions in supervised learning context \cite{fu1998penalized,fan2001variable,armagan2013generalized} and our experiments have shown that the results are not very sensitive to the selection of these hyper-parameters, and thus a default value can be set. On the other hand, the selection of tuning parameter $\lambda$ does have a significant effect on the results, and should be optimized by the grid search.\\

\subsection{Experiments}
\subsubsection{Overfitting of the GSCA model with a fixed rank and no penalty}
The real data sets from the Section 5 are used to show how the GSCA model with a fixed rank and no penalty will overfit the data. The algorithm (details are in the supplementary Section 1) used to fit the GSCA model (with an exact low rank constraint and orthogonality
constraint $\mathbf{A}^{\text{T}}\mathbf{A} = I\mathbf{I}$) is a modification of the developed algorithm in Section 3. GSCA models with three components are fitted using stopping criteria $\epsilon_f = 10^{-5}$ and $\epsilon_f=10^{-8}$. Exactly the same initialization is used for these two models. As shown in Fig.~2, different stopping criteria can greatly effect the estimated $\hat{\mathbf{B}}_1$ from the GSCA models. Furthermore, the number of iterations to reach convergence increases from 141 to 23991. Similar phenomenon, some estimated parameters tend to divergence to plus or negative infinity, has been observed in logistic linear regression model and logistic PCA model \cite{de2006principal, song2017principal}. In logistic linear regression, the estimated coefficients corresponding to the directions where two classes are linearly separable tend to go to plus infinity or minus infinity. The overfitting issue of the GSCA model with exact low rank constraint can be interpreted in the same way by taking the columns of score matrix $\mathbf{A}$ as the latent variables and the loading matrix $\mathbf{B}_1$ as the coefficients to fit the binary $\mathbf{X}_1$. This result suggests that if an exact low rank constraint is preferred in the GSCA model, an extra scale penalty should be added on $\mathbf{B}_1$ to avoid overfitting.\\

\begin{figure}[h!]\label{Fig:2}
    \centering
    \includegraphics[width=\textwidth]{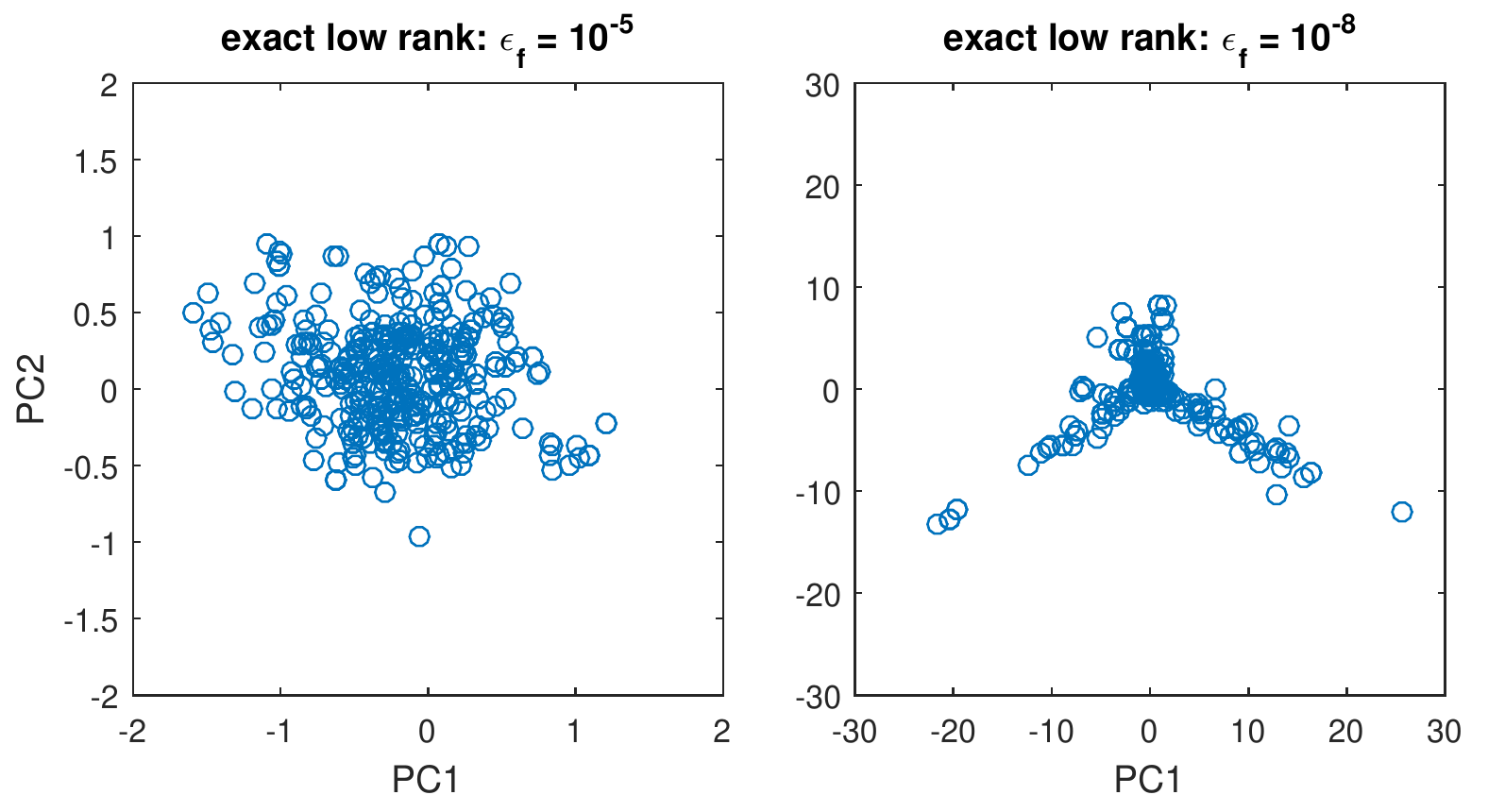}
    \caption*{\textbf{Fig.~2} Loading plots of estimated $\hat{\mathbf{B}}_1$ from the GSCA models with exact low rank constraint using two different stopping criteria $\epsilon_f =10^{-5}$ and $\epsilon_f=10^{-8}$. Note that the scales of the coordinates for $\epsilon_f = 10^{-8}$ (right) is over ten times larger than those for $\epsilon_f = 10^{-5}$ (left).}
\end{figure}

\subsubsection{Comparing the generalization errors of the GSCA models with nuclear norm and concave penalties}
To evaluate the performance of the GSCA model in recovering the underlying structure, we set up the realistic simulation (strongly imbalanced binary data and low SNR) as follows. The simulated $\mathbf{X}_1$ and $\mathbf{X}_2$ have the same size as the real data sets in the Section 5, $I=160$, $J_1=410$, $J_2 = 1000$. The logit transform of the empirical marginal probabilities of the CNA data set in the Section 5 is set as $\bm{\mu}_1$. Elements in $\bm{\mu}_2$ are sampled from the standard normal distribution. The simulated low rank is set to $R=10$; $\sigma^2$ is set to 1; $\text{SNR}_1$ and $\text{SNR}_2$ are set to 1. After the simulation of $\mathbf{X}_1$, there are two columns with identical ``0'' elements, which are removed as they provide no information (no variation).\\

As the GSCA model with the nuclear norm penalty is a convex problem, a global optimum can be obtained. The nuclear norm penalty is therefore used as the baseline in the comparison with other penalties. An interval from $\lambda_0$, which is large enough to achieve an estimated rank of at most rank 1, to $\lambda_{t}$, which is small enough to achieve an estimated rank of 159, is selected based on low precision models ($\epsilon_f=10^{-2}$). 30 log-spaced $\lambda$s are selected equally from the interval $[\lambda_{t},\lambda_{0}]$. The convergence criterion is set as $\epsilon_f = 10^{-8}$. The results are shown in Fig.~3. With decreasing $\lambda$, the estimated rank of $\hat{\mathbf{Z}}$ increased from 0 to 159, and the estimated $\hat{\sigma}^2$ decreased from 2 to close to 0. The minimum $\text{RMSE}(\mathbf{\Theta})$ of 0.184 (the corresponding $\text{RMSE}(\mathbf{\Theta}_1)=0.229$, $\text{RMSE}(\mathbf{\Theta}_2)=0.054$, $\text{RMSE}(\bm{\mu}) = 0.072$ and $\text{RMSE}(\mathbf{Z})=0.446$) can be achieved at $\lambda=38.3$, which corresponds to $\text{rank}(\hat{\mathbf{Z}})=52$ and $\hat{\sigma}^2=0.9271$. There are sharp transitions in all the three subplots near the point $\lambda=40$. The reason is that when the penalty is not large enough, the estimated rank becomes 159, and the constructed GSCA model is almost a saturated model. Thus the model has high generalization error and the estimated $\hat{\sigma}^2$ also becomes close to 0. Given that we only have indirect binary observation $\mathbf{X}_1$ and highly noisy observation $\mathbf{X}_2$ of the underlying structure $\mathbf{\Theta}$, the performance of the GSCA model with nuclear norm penalty is reasonable. However, results can be greatly improved by using concave penalties.\\

\begin{figure}[h!]\label{Fig:3}
    \centering
    \includegraphics[width=\textwidth]{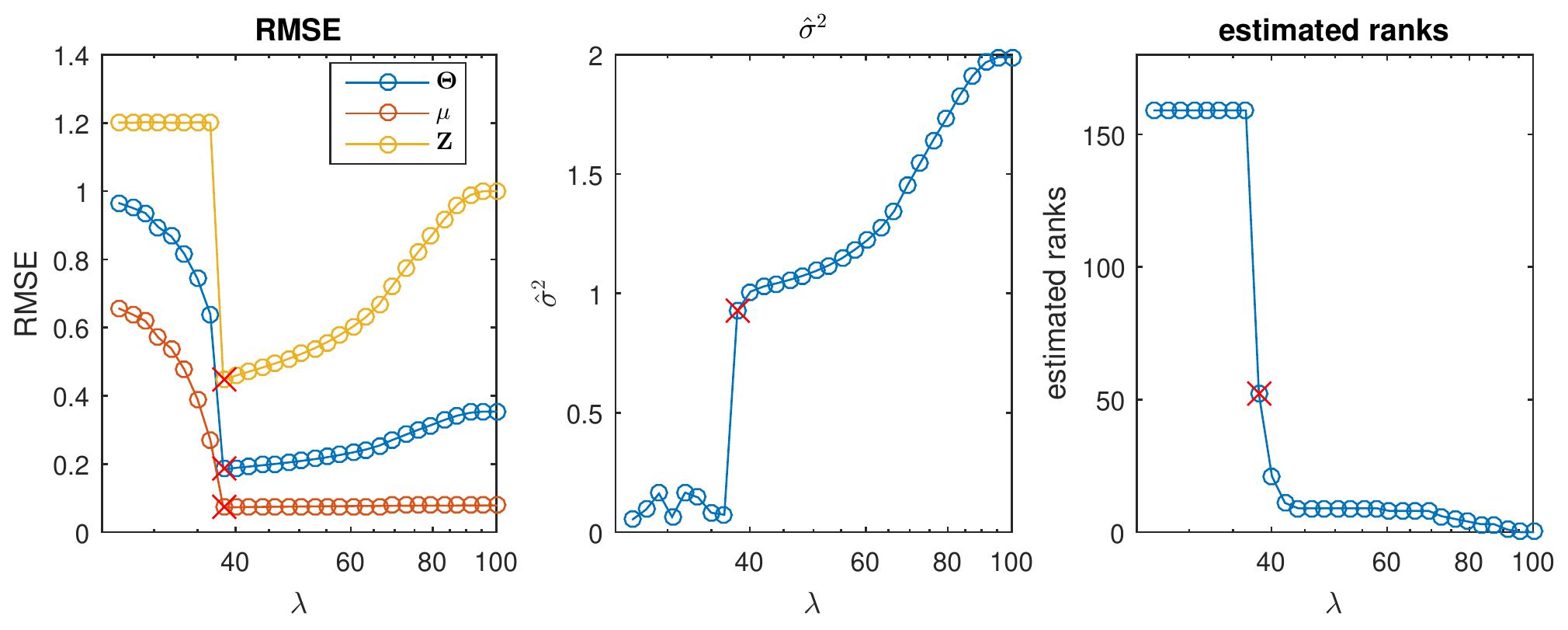}
    \caption*{\textbf{Fig.~3} RMSEs in estimating $\mathbf{\Theta}$, $\bm{\mu}$, $\mathbf{Z}$ (left), the estimated $\hat{\sigma}^2$ (center) and the estimated $\text{rank}(\hat{\mathbf{Z}})$ (right) from the GSCA model with nuclear norm penalty as a function of the tuning parameter $\lambda$. Red cross marker indicates the model with minimum $\text{RMSE}(\mathbf{\Theta})$.}
\end{figure}

For concave penalties, different values of the hyper-parameters, $q$ in $L_q$, $\gamma$ in SCAD and GDP, are selected according to their thresholding properties. For each value of the hyper-parameter, values of tuning parameter $\lambda$ are selected in the same manner as described above. The minimum $\text{RMSE}(\mathbf{\Theta})$ achieved and the corresponding $\text{RMSE}(\bm{\mu})$ and $\text{RMSE}(\mathbf{Z})$ for different values of hyper-parameter of the GSCA models with different penalty functions are shown in Fig.~4. The relationship between RMSEs, $\lambda$ and hyper-parameter for the GSCA model with $L_q$, SCAD and GDP penalty functions are fully characterized in Fig.~S2, Fig.~S3 and Fig.~S4 respectively. As shown in Fig.~4, all GSCA models with concave penalties can achieve much lower RMSEs in estimating $\mathbf{\Theta}$, $\bm{\mu}$ and $\mathbf{Z}$ compared to the convex nuclear norm penalty ($L_{q:q=1}$ in the plot). Among the three concave penalties used, $L_{q}$ and GDP have better performance.\\

\begin{figure}[h!]\label{Fig:4}
    \centering
    \includegraphics[width=\textwidth]{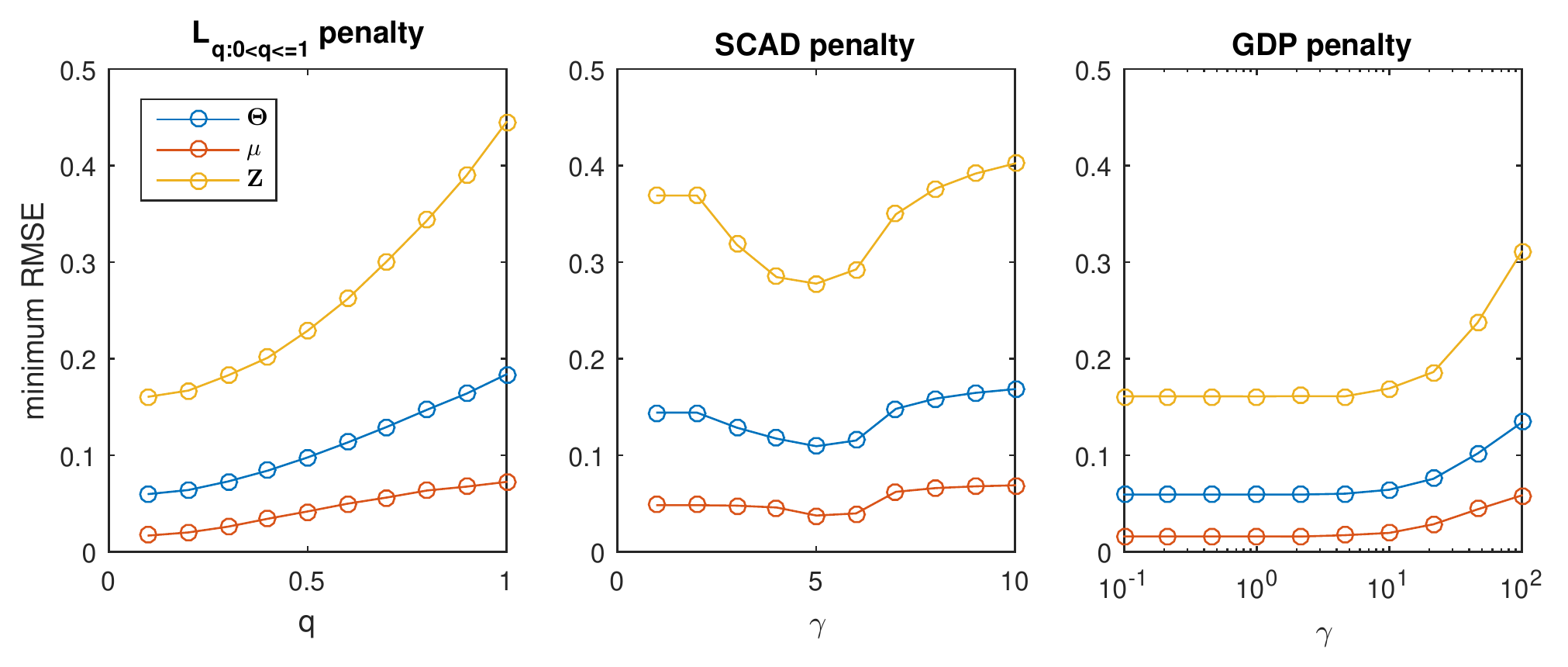}
    \caption*{\textbf{Fig.~4} The minimum $\text{RMSE}(\mathbf{\Theta})$ achieved and the corresponding $\text{RMSE}(\bm{\mu})$ and $\text{RMSE}(\mathbf{Z})$ for different values of hyper-parameter for $L_q$ penalty (left), for SCAD penalty (center) and for GDP penalty (right). The legends indicate the RMSEs in estimating $\mathbf{\Theta}$, $\bm{\mu}$ and $\mathbf{Z}$ respectively. The $x$-axis of the left and center subplots has a linear scale, while right subplot has a log scale.}
\end{figure}

If we get access to the full information, the underlying quantitative data $\mathbf{X}_1^{\ast}$ rather than the binary observation $\mathbf{X}_1$, the SCA model on $\mathbf{X}_1^{\ast}$ and $\mathbf{X}_2$ is simply a PCA model on $[\mathbf{X}_1^{\ast} ~ \mathbf{X}_2]$. From this model, we can get an estimation of $\mathbf{\Theta}$, $\bm{\mu}$ and $\mathbf{Z}$. We compared the results derived from the SCA model on the full information, the GSCA models with nuclear norm, $L_{q:q=0.1}$, SCAD ($\gamma=5$) and GDP ($\gamma=1$) penalties. All the models are selected to achieve the minimum $\text{RMSE}(\mathbf{\Theta})$. The RMSEs of estimating $\mathbf{\Theta}$, $\mathbf{\Theta}_1$, $\mathbf{\Theta}_2$, $\bm{\mu}$ and $\mathbf{Z}$ and the rank of estimated $\hat{\mathbf{Z}}$ from different models are shown in Table 2. Here we can see that the GSCA models with $L_{q:q=0.1}$ and GDP ($\gamma=1$) penalties have better performance in almost all criteria compared to the nuclear norm and SCAD penalties, and even comparable with the SCA model on full information. The singular values of the true $\mathbf{Z}$, estimated $\hat{\mathbf{Z}}$ from the above models and the noise terms $\mathbf{E} = [\mathbf{E}_1~\mathbf{E}_2]$ are shown in Fig.~5. Only the first 15 singular values are shown to have higher resolution of the details. Since the $10$-th singular value of the simulated data $\mathbf{Z}$ is smaller than the noise level, the best achievable rank estimation is 9. Both the $L_{q:q=0.1}$ and GDP ($\gamma=1$) penalties successfully find the correct rank 9, and they have a very good approximation of the first 9 singular values of $\mathbf{Z}$. On the other hand, the nuclear norm penalty shrinks all the singular values too much. Furthermore, the SCAD penalty overestimates the first three singular values and therefore shrinks all the other singular values too much. These results are easily understandable if taking their thresholding properties in Fig.~2 into account. Both the $L_{q}$ and the GDP penalties have very good performance in this simulation experiment.\\

\begin{table}[htbp]
\centering
\caption*{\label{tabS1} Table 2: The RMSEs of estimating $\mathbf{\Theta}$, $\bm{\mu}$ and $\mathbf{Z}$ and the rank of estimated $\hat{\mathbf{Z}}$ from different models.}
\begin{tabular}{|l|l|l|l|l|l|l|}
 \hline
    & $\text{RMSE}(\mathbf{\Theta})$ & $\text{RMSE}(\mathbf{\Theta}_1)$ & $\text{RMSE}(\mathbf{\Theta}_2)$ & $\text{RMSE}(\bm{\mu})$ & $\text{RMSE}(\mathbf{Z})$ & $\text{rank}(\hat{\mathbf{Z}})$\\
 \hline
  $L_{q:q=1}$   & 0.1840 & 0.2288 & 0.0537 & 0.0724 & 0.4456 & 52 \\
  $L_{q:q=0.1}$ & 0.0598 & 0.0682 & 0.0353 & 0.0168 & 0.1606 & 9\\
  SCAD($\gamma=5$) & 0.1093 & 0.1334 & 0.0395 & 0.0376 & 0.2777 & 24 \\
  GDP($\gamma=1$) & 0.0593 & 0.0675 & 0.0354 & 0.0160 & 0.1610 & 9 \\
  full information & 0.0222 & 0.0675 & 0.0354 & 0.0030 & 0.0674 & 9 \\
  \hline
\end{tabular}
\end{table}

\begin{figure}[h!]\label{Fig:5}
    \centering
    \includegraphics[width=0.5 \textwidth]{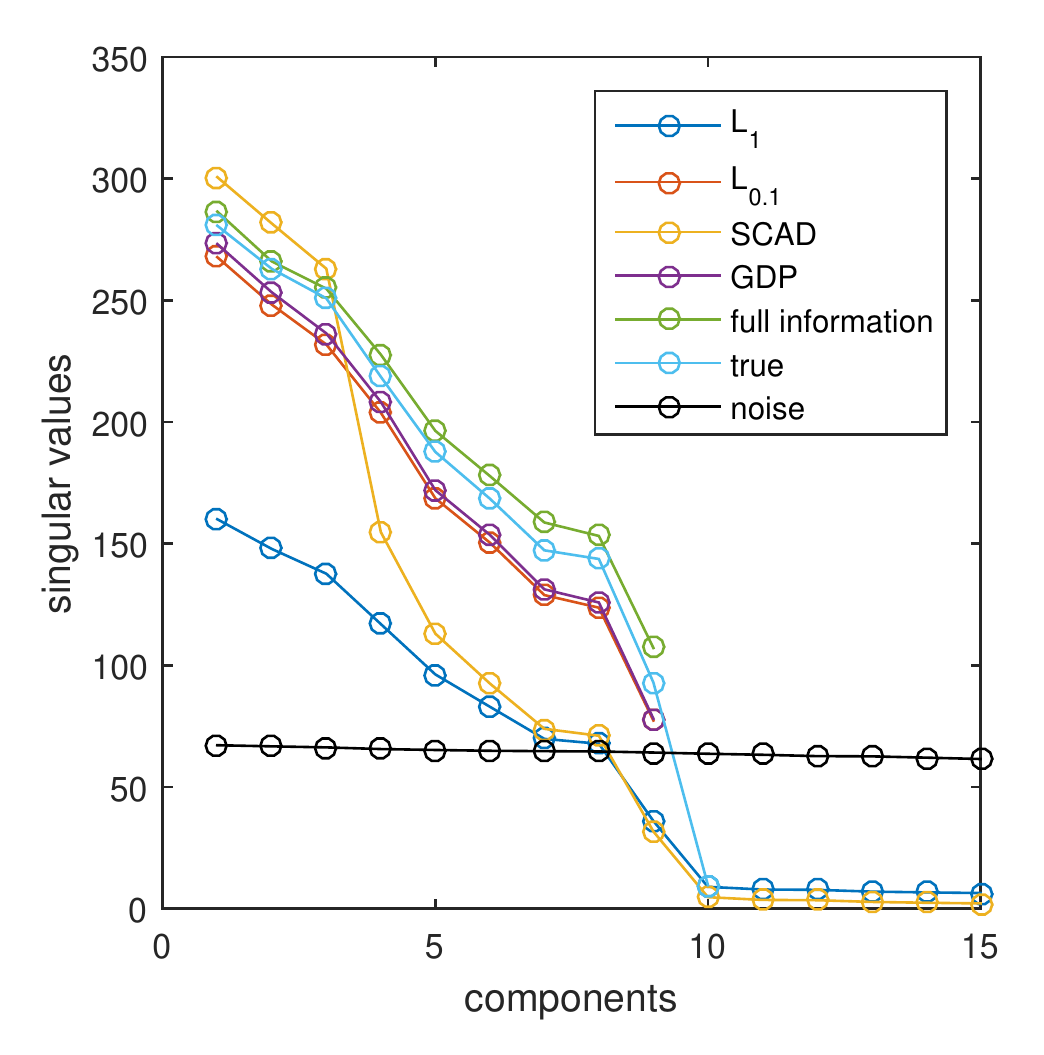}
    \caption*{\textbf{Fig.~5} Approximation of the singular values using different penalties in the simulation experiment. Labels ``$L_1$'', ``$L_{0.1}$'', ``SCAD'', ``GDP'', ``full information'' indicate the singular values of estimated $\hat{\mathbf{Z}}$ from the corresponding models; ``$\text{true}$'' indicates the singular values of the simulated $\mathbf{Z}$; ``$\text{noise}$'' indicates the singular values of the noise term $\mathbf{E}$, which has full rank.}
\end{figure}

\subsubsection{Comparing the GSCA model with GDP penalty and the iClusterPlus model}
A detailed theoretical comparison of our method to the iClusterPlus model \cite{mo2013pattern} and a related work \cite{wu2015fast} can be found in the supplementary Section 3. After that, we compared our GSCA model with GDP penalty to the iClusterPlus model on the simulated data sets. The parameters for the GSCA model with GDP penalty is the same as described above. The running time is 60.61s when $\epsilon_f=10^{-8}$, and 9.98s when $\epsilon_f=10^{-5}$. For the iClusterPlus model, 9 latent variables are specified. The tuning parameter of the lasso type constraint on the data specific coefficient matrices are set to 0. The default convergence criterion is used, that is the maximum of the absolute changes of the estimated parameters in two subsequent iterations is less than $10^{-4}$. The running time of the iClusterPlus model is close to 3 hours. The constructed iClusterPlus model provides the estimation of column offset $\hat{\bm{\mu}}$, the common latent variables $\hat{\mathbf{A}}$, and data set specific coefficient matrices $\hat{\mathbf{B}}_1$ and $\hat{\mathbf{B}}_2$. The estimated $\hat{\mathbf{Z}}$ and $\hat{\mathbf{\Theta}}$ are computed in the same way as defined in the model section. The RMSEs in estimating $\mathbf{\Theta}$, $\bm{\mu}$ and $\mathbf{Z}$ for iClusterPlus are 2.571, 2.473 and 3.060 respectively. Compared to the results from the GSCA models in Table 2, iClusterPlus is unable to provide good results on the simulated data sets. Fig.~S5 compares the estimated $\hat{\mu}_1$ from the GSCA model with GDP penalty and iClusterPlus model. As shown in Fig.~S5(right), the iClusterPlus model is unable to estimate the offset $\bm{\mu}$ correctly. Many elements of estimated $\hat{\bm{\mu}_1}$ are exactly 0, which corresponds to an estimated marginal probability of 0.5. In addition, as shown in Fig.~6(left), the singular values of the estimated $\hat{\mathbf{Z}}$ from the iClusterPlus model are clearly overestimated. These undesired results from the iClusterPlus model are due mainly to the imbalancedness of the simulated binary data set. If the offset term $\bm{\mu}_1$ in the simulation is set to 0, which corresponds to balanced binary data simulation, and fix all the other parameters in the same way as in the above simulation, the results of iClusterPlus and the GSCA with GDP penalty are more comparable. In that case the RMSEs of estimating $\mathbf{\Theta}$, $\mathbf{Z}$ in the GSCA model with GDP penalty are 0.071 and 0.091 respectively, while the RMSEs of the iClusterPlus model are 0.107 and 0.142 respectively. As shown in Fig.~6(right), the singular values of estimated $\hat{\mathbf{Z}}$ from the iClusterPlus model are much more accurate compared to the imbalanced case. However, iClusterPlus still overestimates the singular values compared to the GSCA model with GDP penalty. This phenomenon is related to the fact that exact low rank constraint is also used in the iClusterPlus model. These results suggest that compared to iClusterPlus, the GSCA model with GDP penalty is more robust to the imbalanced binary data and has better performance in recovering the underlying structure in the simulation experiment.\\

\begin{figure}[h!]\label{Fig:6}
    \centering
    \includegraphics[width=\textwidth]{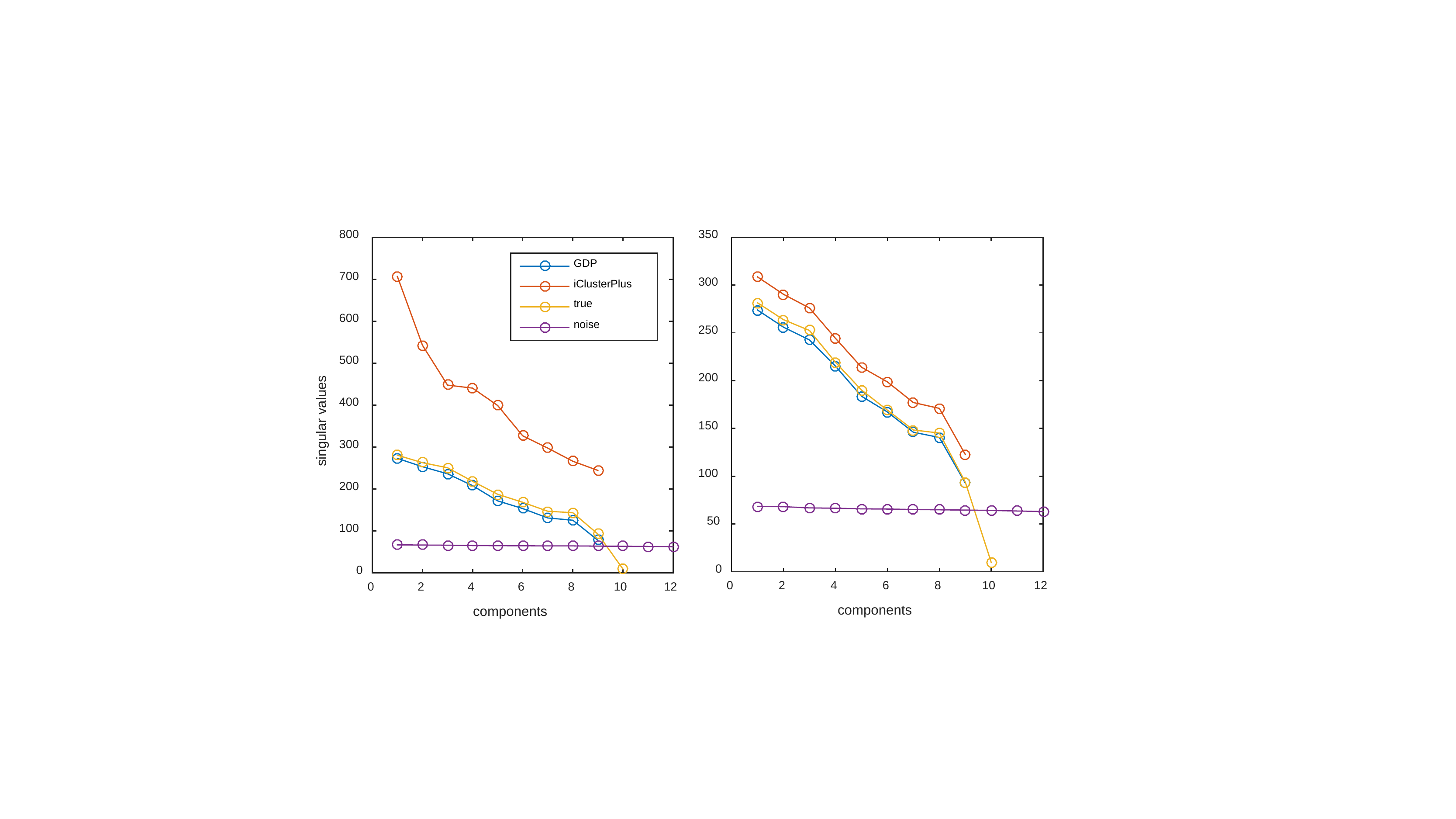}
    \caption*{\textbf{Fig.~6} The singular values of estimated $\hat{\mathbf{Z}}$ using the iClusterPlus model and the GSCA model with GDP penalty on the simulation with imbalanced binary data (left) and with balanced binary data (right).}
\end{figure}

\subsubsection{The performance of the GSCA model for the simulation with different SNRs}
We will explore the performance of the GSCA model for the simulated binary and quantitative data sets with varying noise levels in the following experiment. Equal SNR levels are used in the simulation for $\mathbf{X}_1$ and $\mathbf{X}_2$. 20 log spaced SNR values are equally selected from the interval $[0.1, 100]$. Then we simulated coupled binary data $\mathbf{X}_1$ and quantitative $\mathbf{X}_2$ using the different SNRs in the same way as described above. During this process, except for the parameters $c_1$ and $c_2$, which are used to adjust the SNRs, all other parameters used in the simulation were kept the same. The GSCA models with GDP penalty ($\gamma=1$), $L_{q}$ penalty ($q=0.1$), nuclear norm penalty, and the SCA model on the full information (defined above) are used in these simulation experiments. For these three models, the model selection process was done in the same way as described in above experiment. The models with the minimum $\text{RMSE}(\mathbf{\Theta})$ are selected. As shown in Fig.~7, the GSCA models with concave GDP and $L_{q}$ penalties always have better performance than the convex nuclear norm penalty, and they are comparable to the situation where the full information is available. With the increase of SNR, the $\text{RMSE}(\mathbf{Z})$ derived from the GSCA model, which is used to evaluate the performance of the model in recovering the underlying low dimensional structure, first decreases to a minimum and then increases. As shown in bottom center and right, this pattern is mainly caused by how $\text{RMSE}(\mathbf{Z}_1)$ changes with respect to SNRs. Although this result counteracts the intuition that larger SNR means higher quality of data, it is in line with previous results on logistic PCA model of binary data set \cite{davenport20141}. In order to understand this effect, considering the S-shaped logistic curve, the plot of the function $\phi(\theta) = (1+\exp(-\theta))^{-1}$. This curve almost becomes flat when $\theta$ becomes very large. There is no resolution anymore in these flat regimes. A large deviation in $\theta$ has almost no effect on the logistic response. When the SNR becomes extremely large, the scale of the simulated parameter $\theta$ is very extreme, then even if we have a good estimation of the probability $\hat{\pi} = \phi(\hat{\theta})$, the scale of estimated $\hat{\theta}$ can be far away from the simulated $\theta$. We refer \cite{davenport20141} for a detailed interpretation of this phenomenon.\\

\begin{figure}[h!]\label{Fig:7}
    \centering
    \includegraphics[width=\textwidth]{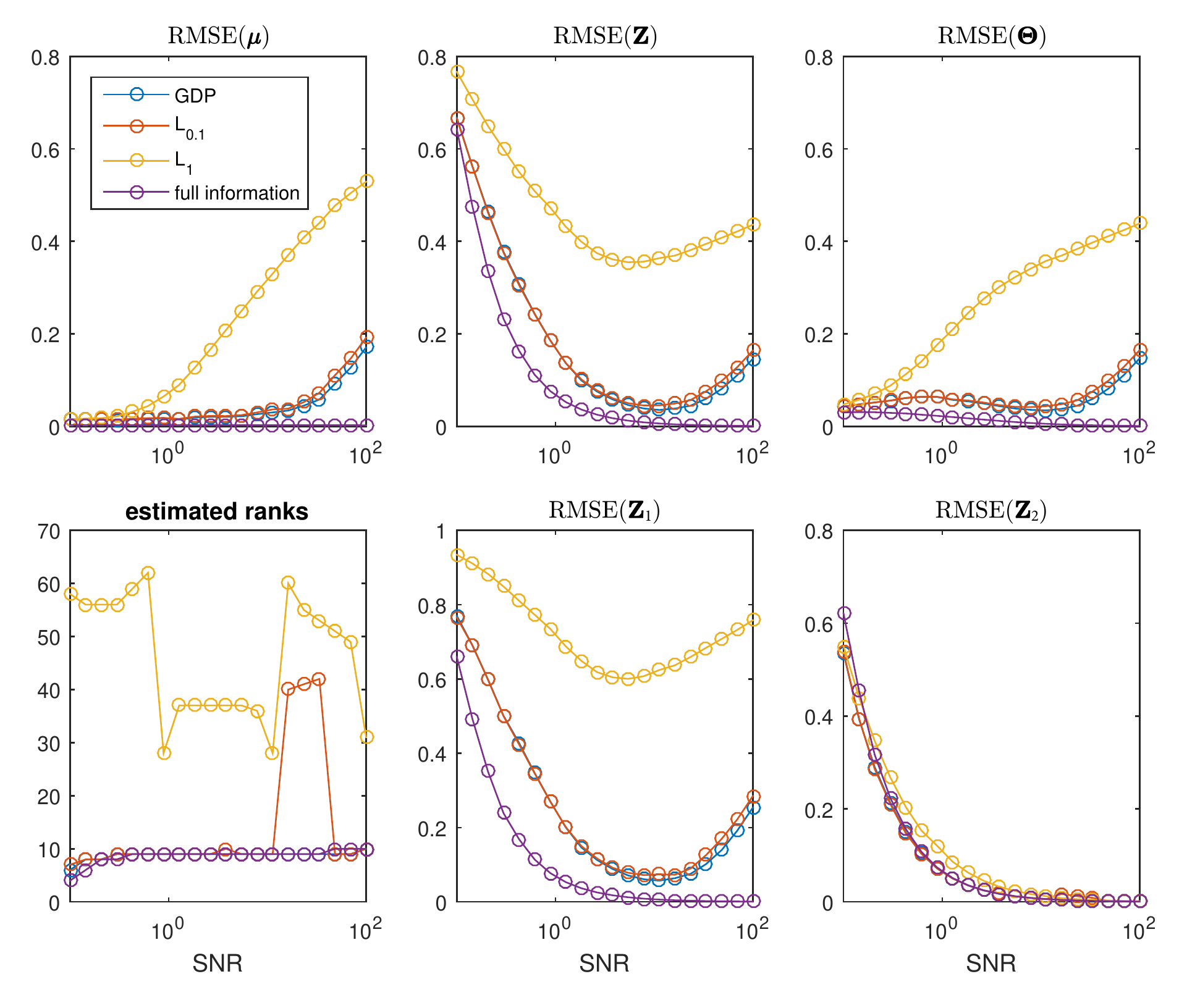}
    \caption*{\textbf{Fig.~7} Minimum $\text{RMSE}(\mathbf{\Theta})$ (top right), and the corresponding $\text{RMSE}(\bm{\mu})$ (top left), $\text{RMSE}(\mathbf{Z})$ (top center), rank estimation of $\hat{\mathbf{Z}}$ (bottom left), $\text{RMSE}(\mathbf{Z}_1)$ (bottom center) and $\text{RMSE}(\mathbf{Z}_2)$ (bottom right) of the GSCA models with nuclear norm penalty (legend ``$L_{1}$''), GDP penalty (legend GDP), $L_{0.1}$ penalty (legend ``$L_{0.1}$'') and SCA model on full information (legend ``full information'') for different SNR levels.}
\end{figure}

\subsubsection{Assessing the model selection procedure}
The cross validation procedure and the cross validation error have been defined in the model selection section. The GSCA model with GDP penalty is used as an example to assess the model selection procedure. $\epsilon_f=10^{-5}$ is used as the stopping criteria for all the following experiments to save time. The values of $\lambda$ and $\gamma$ are selected in the same way as was described in Section 4.2. Fig.~8 shows the minimum $\text{RMSE}(\mathbf{\Theta})$ and minimum CV error achieved for different values of the hyper-parameter $\gamma$. The minimum CV error changes in a similar way as the minimum $\text{RMSE}(\mathbf{\Theta})$ with respect to the values of $\gamma$. However, taking into account the uncertainty of estimated CV errors, the difference of the minimum CV errors for different $\gamma$ is very small. Thus, we recommend to fix $\gamma$ to be 1, rather than using cross validation to select it. Furthermore, setting $\gamma = 1$ as the default value for the GDP penalty has a probabilistic interpretation, see in \cite{armagan2013generalized}.\\

\begin{figure}[h!]\label{Fig:8}
    \centering
    \includegraphics[width=\textwidth]{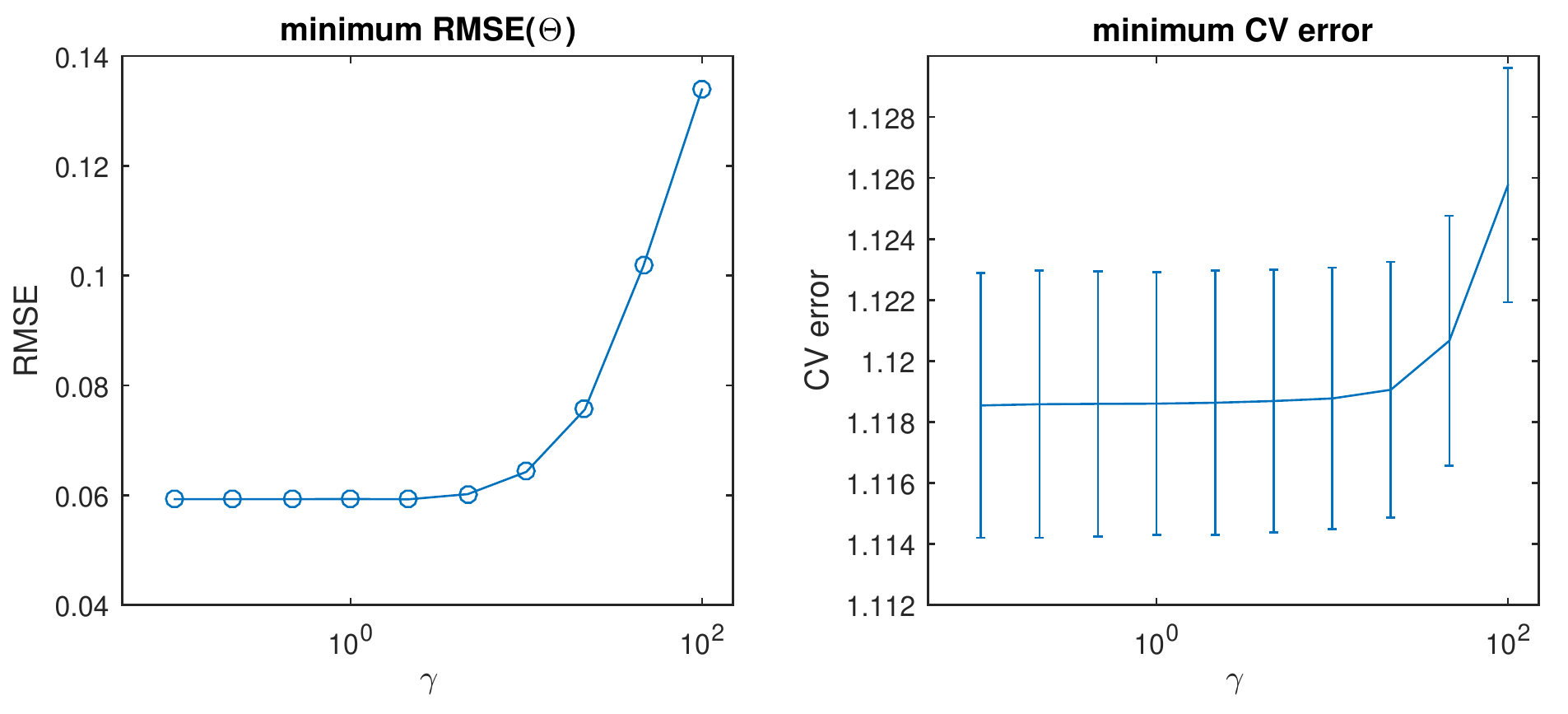}
    \caption*{\textbf{Fig.~8} Minimum $\text{RMSE}(\mathbf{\Theta})$ (left) and minimum CV error (right) for different values of $\gamma$ from the GSCA model with GDP penalty. One standard error bars are added to the CV error plot.}
\end{figure}

Whenever the GSCA model is used for exploratory data analysis, there is no need to select $\lambda$ explicitly. It is sufficient to find a proper value to achieve a two or three component GSCA model, in order to visualize the estimated score and loading matrices. If the goal is confirmatory data analysis, it is possible to select the tuning parameter $\lambda$ explicitly by the proposed cross validation procedure. Fig.~9 shows how the tuning parameter $\lambda$ affects the CV errors, $\text{RMSE}(\mathbf{\Theta})$ and the estimated ranks. The minimum CV error obtained is close to the Bayes error, which is the scaled negative log likelihood in cases where the true parameters $\mathbf{\Theta}$ and $\sigma^2$ are known. Even through, inconsistence exists between CV error plot (Fig.~9, left) and the $\text{RMSE}(\mathbf{\Theta})$ plot (Fig.~9, center), the selected model corresponding to minimum CV error can achieve very low $\text{RMSE}(\mathbf{\Theta})$ and correct rank estimation (Fig.~9, right). Therefore, we suggest to use the proposed CV procedure to select the value of $\lambda$ at which the minimum CV error is obtained. Finally, we fit a model on full data set without missing elements using the selected value of $\lambda$ and the outputs of the selected model with minimum CV error as the initialization.\\

\begin{figure}[h!]\label{Fig:9}
    \centering
    \includegraphics[width=\textwidth]{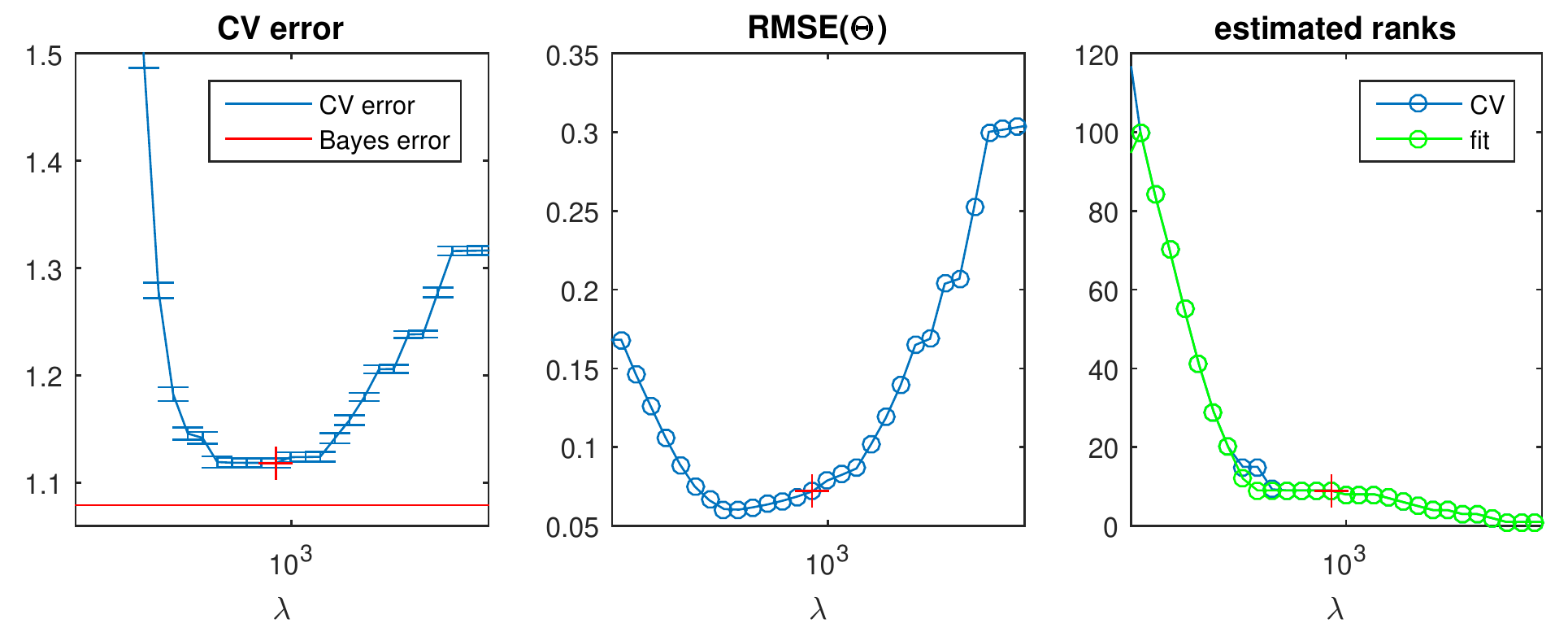}
    \caption*{\textbf{Fig.~9} CV error, RMSE and estimated rank for different values of the tuning parameter $\lambda$. One standard error bars are added to the CV error plot. ``Bayes error'' indicates the mean log negative likelihood using simulated $\mathbf{\Theta}$ and $\sigma^2$ to fit the simulated data sets $\mathbf{X}_1$ and $\mathbf{X}_2$. The red cross marker indicates the point where the minimum CV error is achieved. ``CV'' and ``fit'' (right plot) indicate the mean $\text{rank}(\hat{\mathbf{Z}})$ derived from the models constructed in 7-fold cross validation and the $\text{rank}(\hat{\mathbf{Z}})$ derived from a model constructed on full data set without missing elements (the outputs of the constructed model during cross validation are set as the initialization.}
\end{figure}

\section{Empirical illustration}
\subsection{Real data set}
The Genomic Determinants of Sensitivity in Cancer 1000 (GDSC1000) \cite{iorio2016landscape} contains 926 tumor cell lines with comprehensive measurements of point mutation, CNA, methylation and gene expression. We selected the binary CNA and quantitative gene expression measurements on the same cell lines (each cell line is a sample) as an example to demonstrate the GSCA model. To simplify the interpretation of the derived model, only the cell lines of three cancer types are included: BRCA (breast invasive carcinoma, 48 cell lines), LUAD (lung adenocarcinoma, 62 cell lines) and SKCM (skin cutaneous melanoma, 50 cell lines). The CNA data set has 410 binary variables. Each variable is a copy number region, in which ``1'' indicates the presence and ``0'' indicates the absence of an aberration. Note that, the CNA data is very imbalanced: only $6.66\%$ the elements are ``1''. The empirical marginal probabilities of binary CNA variables are shown in Fig.~S1. The quantitative gene expression data set contains 17,420 variables, of which 1000 gene expression variables with the largest variance are selected. After that, the gene expression data is column centered and scaled by the standard deviation of the each variables to make it more consistent with the assumption of the GSCA model.\\

\subsection{Exploratory data analysis of the coupled CNA and gene expression data sets}
We applied the GSCA model (with GDP penalty and $\gamma$=1) to the GDSC data set of 160 tumor cell lines that have been profiled for both binary CNA ($160 \times 410$) and quantitative gene expression ($160 \times 1000$). The results of model selection (Fig.~S6) validate the existence of a low dimensional common structure between CNA and gene expression data sets. For exploratory purposes, we will construct a three component model instead.\\

We first considered the score plot resulting from this GSCA model. The first two PCs show a clear clustering by cancer type (Fig.~10, left), and in some cases even subclusters (i.e. hormone-positive breast cancer, MITF-high melanoma). These results suggest that the GSCA model captures the relevant biology in these data. Interestingly, when we performed PCA on the gene expression data, we obtained score plots that were virtually identical to those resulting from the GSCA model (Fig.~S7, left; modified RV coefficient: 0.9998), suggesting that this biological relevance is almost entirely derived from the gene expression data.\\

\begin{figure}[h!]\label{Fig:10}
    \centering
    \includegraphics[width=\textwidth]{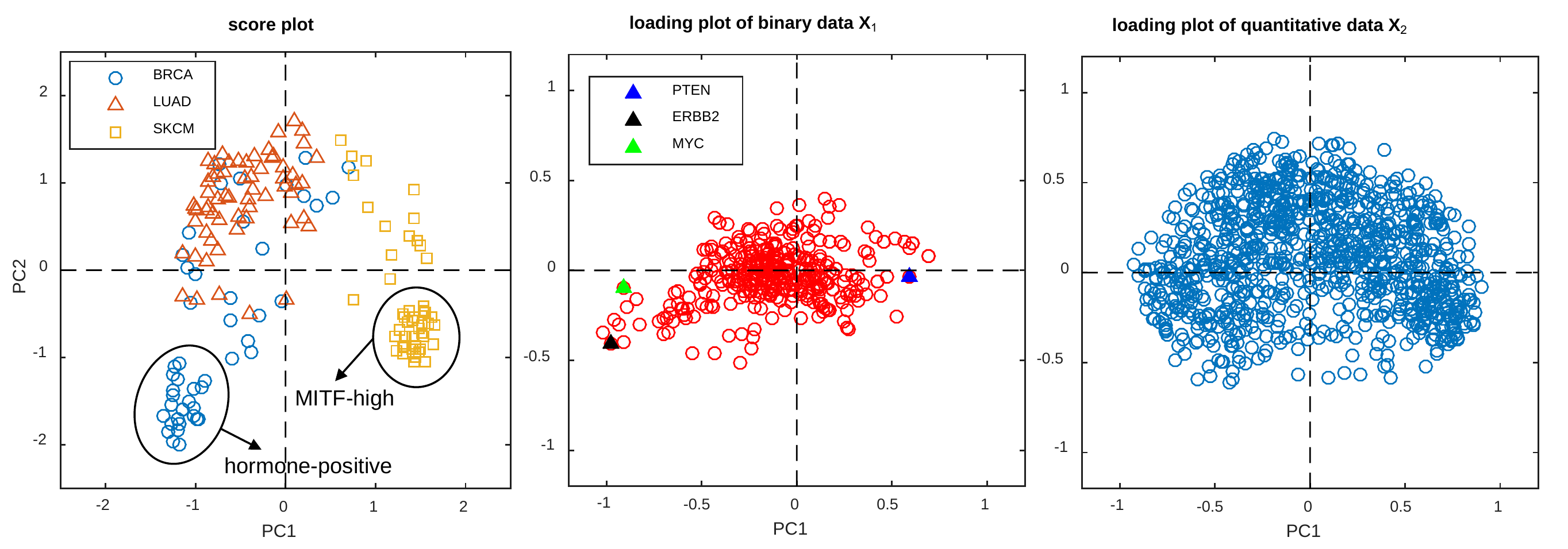}
    \caption*{\textbf{Fig.~10} Score plot (left), loading plot for binary CNA data $\mathbf{X}_1$ (center) and loading plot for gene expression data $\mathbf{X}_2$ (right) derived from the constructed GSCA model.}
\end{figure}

We then wondered whether the GSCA model could leverage the gene expression data to help us gain insight into the CNA data. To test this, we first established how much insight could be gained from the CNA data in isolation. Fig.~S8 shows the scores and loadings of the first two components from a three component logistic PCA model \cite{de2006principal} applied to the CNA data. While these do seem to contain structure in the loading plot, we believe that they mostly explain technical characteristics of the data. For example, deletions and amplifications are almost perfectly separated from each other by the PC1=0 line in the loading plot (Fig.~S9). Additionally, the scores on PC1 are strongly associated to the number of copy number aberrations (i.e., to the number of ones) in a given sample (Fig.~S10). Finally, the clusters towards the left of the loading plot suggested two groups of correlated features, but these could trivially be explained by genomic position, that is, these features correspond with regions on the same chromosome arm, which are often completely deleted or amplified (Fig.~S11). Following these observations, we believe that a study of the CNA data in isolation provides little biological insight.\\

On the other hand, using the GCSA model’s CNA loadings (Fig.~10, center), we could more easily relate the features to the biology. Let us focus on features with extreme values on PC1 and for which the corresponding chromosomal region contains a known driver gene. For example, the position of MYC amplifications in the loading plot indicates that MYC amplifications occur mostly in lung adenocarcinoma and breast cancer samples (Fig.~10, center; Fig.~S12). Similarly, ERBB2 amplifications occur mainly in breast cancer samples (Fig.~10, center; Fig.~S12). Finally, PTEN deletions were enriched in melanomas, though the limited size of the loading also indicates that they are not exclusive to melanomas (Fig.~10, center; Fig.~S12). Importantly, these three findings are in line with known biology \cite{akbani2015genomic,cancer2014comprehensive,cancer2012comprehensive} and hence exemplify how GSCA could be used to interpret the CNA data. Altogether, using the GSCA model, we were able to 1) capture the biological relevance in the gene expression data, and 2) leverage that biological relevance from the gene expression to gain a better understanding of the CNA data.\\

\section{Discussion}
In this paper, we generalized the standard SCA model to explore the dependence between coupled binary and quantitative data sets. However, the GSCA model with exact low rank constraint overfits the data, as some estimated parameters tend to divergence to positive infinity or negative infinity. Therefore, concave penalties are introduced in the low rank approximation framework to achieve low rank approximation and to mitigate the overfitting issues of the GSCA model. An efficient algorithm framework with analytical form updates for all the parameters is developed to optimize the GSCA model with any concave penalties. All concave penalties used in our experiments have better performance with respect to generalization error and estimated low rank of the constructed GSCA model compared to the nuclear norm penalty. Both $L_{q}$ and GDP penalties with proper model selection can recover the simulated low rank structures almost exactly only from indirect binary observation $\mathbf{X}_1$ and noisy quantitative observation $\mathbf{X}_2$. Furthermore, we have shown that the GSCA model outperforms the iClusterPlus model with respect to speed and accuracy of the estimation of the model parameters.\\

Why the GSCA models with concave penalties have better performance? The exact low rank constraint thresholds the singular values in a hard manner and, therefore, only the largest $R$ singular values are kept. On the other hand, the nuclear norm penalty works in a soft manner, in which all the singular values are shrunk by the same amount of $\lambda$. The thresholding properties of the concave penalties discussed in this paper lie in between these two approaches. As $\mathbf{Z}=\mathbf{A}\mathbf{B}^{\text{T}}$ and $\mathbf{A}^{\text{T}}\mathbf{A}= I\mathbf{I}_R$, the scale of the loadings is related to the scale of the singular values of $\mathbf{Z}$. Thus, we can shrink the singular values of $\mathbf{Z}$ to control the scale of estimated loading matrices in an indirect way. The exact low rank constraint kept the $R$ largest singular values but without control of the scale of the estimated singular values, leading to overfitting. On the other hand, nuclear norm penalty shrinks all the singular values by the same amount of $\lambda$, leading to biased estimation of the singular values. A concave penalty, like $L_q$ or GDP, achieves a balance in thresholding the singular values. Among the concave penalties we used in the experiment, the SCAD penalty does not work well in the simulation study. The reason is that the SCAD penalty does not shrink the large singular values, which therefore tend to be overfitted, while the smaller singular values are shrunk too much.\\

Compared to the iClusterPlus method, only the option of binary and quantitative data sets are included in our GSCA model, and at the moment no sparsity can be imposed for the integrative analysis of binary and quantitative data sets. However, the GSCA model with GDP penalty is optimized by a more efficient algorithm, it is much more robust to the imbalanced nature of the biological binary data and it provides a much better performance for the simulation experiments in this paper. Furthermore, the exploratory analysis of the GDSC coupled CNA and gene expression data sets provided important information on the binary CNA data that was not obtained by a separate analysis.\\

\bibliographystyle{ieeetr}

\bibliography{reference}

\clearpage
\section*{Supplementary material}
\subsection*{GSCA model with exact low rank constraint}
The exact low rank constraint on $\mathbf{Z}$ can be expressed as the multiplication of two low rank matrices $\mathbf{A}$ and $\mathbf{B}$. The optimization problem related to the GSCA model with exact low rank constraint can be expressed as
\begin{equation}
\begin{aligned}
    \min_{\bm{\mu},\mathbf{Z},\sigma^2} \quad & f_1(\mathbf{\Theta}_1) + f_2(\mathbf{\Theta}_2,\sigma^2) \\
    \text{s.t.~} \mathbf{\Theta} &= \mathbf{1}\bm{\mu}^{\text{T}} + \mathbf{Z} \\
     \mathbf{\Theta} &= [\mathbf{\Theta}_1 ~ \mathbf{\Theta}_2] \\
    \text{rank}(\mathbf{Z}) &= R
\end{aligned}
\end{equation}

The developed algorithm in the paper can be slightly modified to fit this model. Same as in the paper, the above optimization problem can majorized to the following problem.
\begin{equation}
\begin{aligned}
              & \frac{L}{2} ||\mathbf{\Theta} - \mathbf{H}^k||_F^2+ c\\
        \mathbf{\Theta} &= \mathbf{1}\bm{\mu}^{\text{T}} + \mathbf{Z} \\
         \mathbf{H}^k &= \mathbf{\Theta}^k - \frac{1}{L} (\mathbf{Q} \odot \nabla f(\mathbf{\Theta}^k))\\
            \mathbf{1}^{\text{T}}\mathbf{Z} &= 0 \\
             \text{rank}(\mathbf{Z}) &= R.
\end{aligned}
\end{equation}

The analytical solution of the $\bm{\mu}$ is also the column mean of $\mathbf{H}^k$. After deflating out the offset term $\bm{\mu}$, the majorized problem becomes $\min_{\mathbf{Z}} \frac{L}{2} ||\mathbf{Z} - \mathbf{J} \mathbf{H}^k||_F^2 \quad  \text{s.t.} \quad  \text{rank}(\mathbf{Z}) = R$, $\mathbf{1}^{\text{T}}\mathbf{Z} = 0$. The global optimal solution is the $R$ truncated SVD of $\mathbf{J} \mathbf{H}^k$. Other steps in the algorithm to fit the GSCA model with exact low rank constraint are exactly the same the algorithm developed in the paper to fit the GSCA model with concave penalties.\\

\subsection*{Figures and tables}

\begin{table}[htbp]
\centering
\caption*{\label{tabS1} Table S1: Comparison of the average computational time (in seconds) of the GSCA model with different penalties and the corresponding 7-fold CV procedure. The binary CNA and quantitative gene expression data sets are used as an example. ``fit'': a three components GSCA model; ``CV'': 7-fold CV procedure. All the models are repeated 5 times, the average computational time is recorded.}
\begin{tabular}{|l|l|l|l|}
 \hline
       & fit: $\epsilon_f=10^{-5}$ & CV: $\epsilon_f=10^{-5}$ & fit: $\epsilon_f=10^{-8}$\\
  \hline
  $L_{1}$   & 9.68  & 18.33 & 57.48\\
  $L_{0.1}$ & 11.28 & 25.06 & 67.47\\
  SCAD($\gamma=5$) & 9.96  & 18.44  & 57.58 \\
  GDP($\gamma=1$)  & 11.90 & 27.18  & 69.66 \\
  \hline
\end{tabular}
\end{table}

\begin{figure}[h!]\label{Fig:S1}
    \centering
    \includegraphics[ width= 0.5\textwidth]{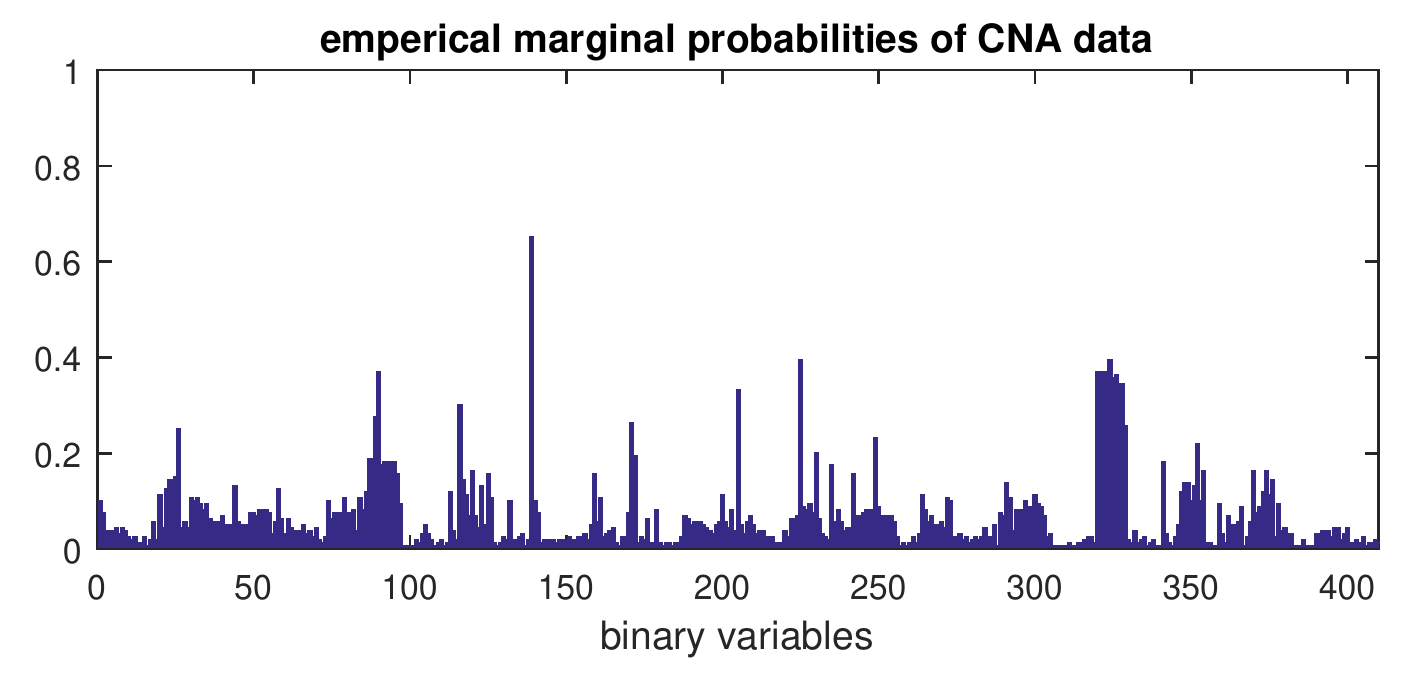}
    \caption*{\textbf{Fig.~S1} Empirical marginal probabilities of binary CNA data set.}
\end{figure}

\begin{figure}[h!]\label{Fig:S2}
    \centering
    \includegraphics[width=\textwidth]{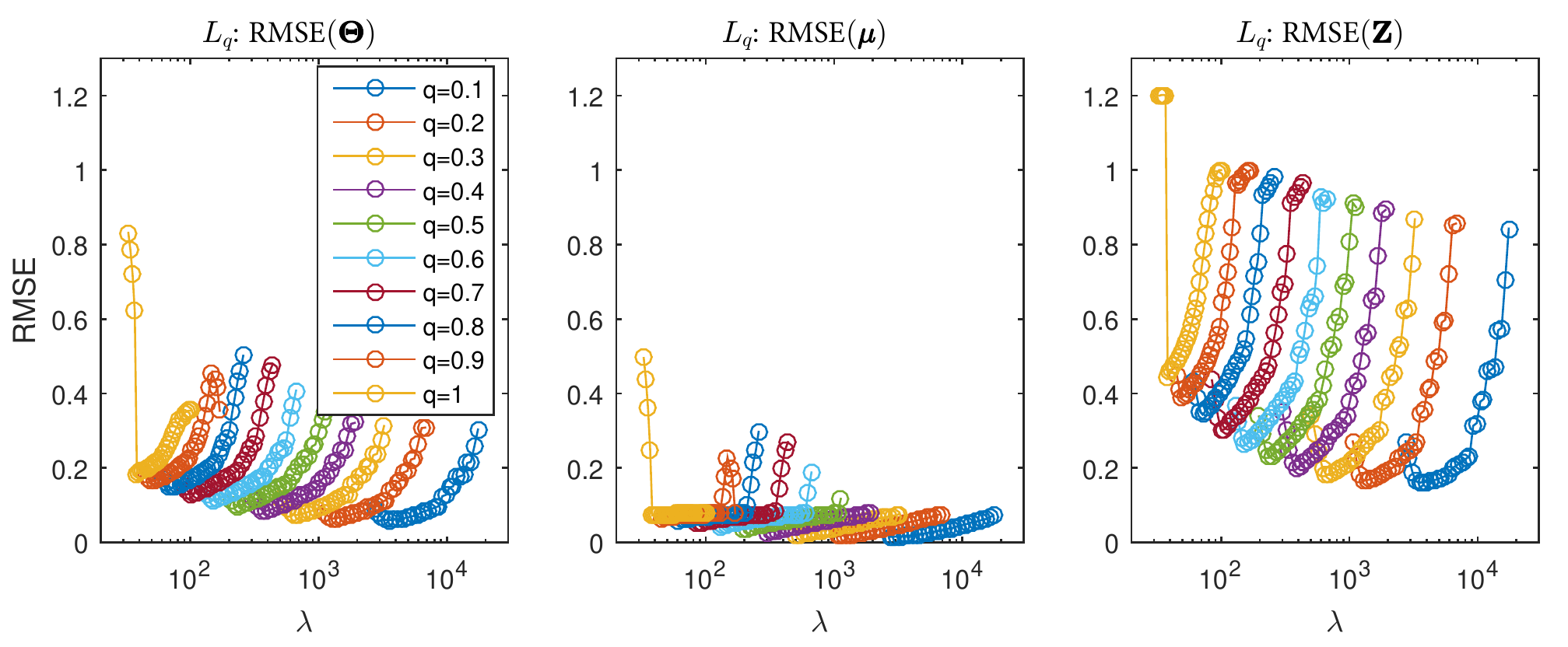}
    \caption*{\textbf{Fig.~S2} Relationship between $q$, $\lambda$ and RMSEs of estimating $\mathbf{\Theta}$, $\bm{\mu}$ and $\mathbf{Z}$ achieved for the GSCA model with $L_q$ penalty. The $x$-axis has a log scale.}
\end{figure}

\begin{figure}[h!]\label{Fig:S3}
    \centering
    \includegraphics[width=\textwidth]{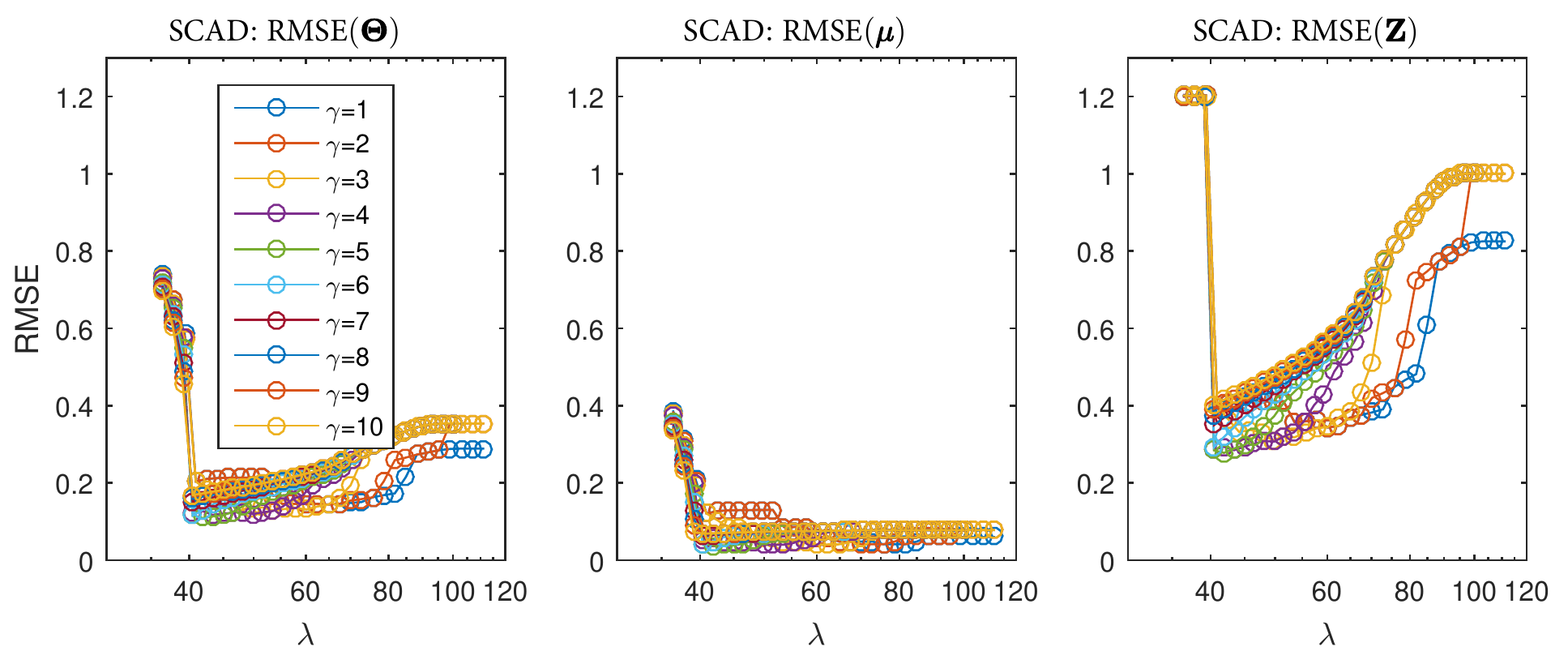}
    \caption*{\textbf{Fig.~S3} Relationship between $\gamma$, $\lambda$ and RMSEs of estimating $\mathbf{\Theta}$, $\bm{\mu}$ and $\mathbf{Z}$ achieved for the GSCA model with SCAD penalty. The $x$-axis has a log scale.}
\end{figure}

\begin{figure}[h!]\label{Fig:S4}
    \centering
    \includegraphics[width=\textwidth]{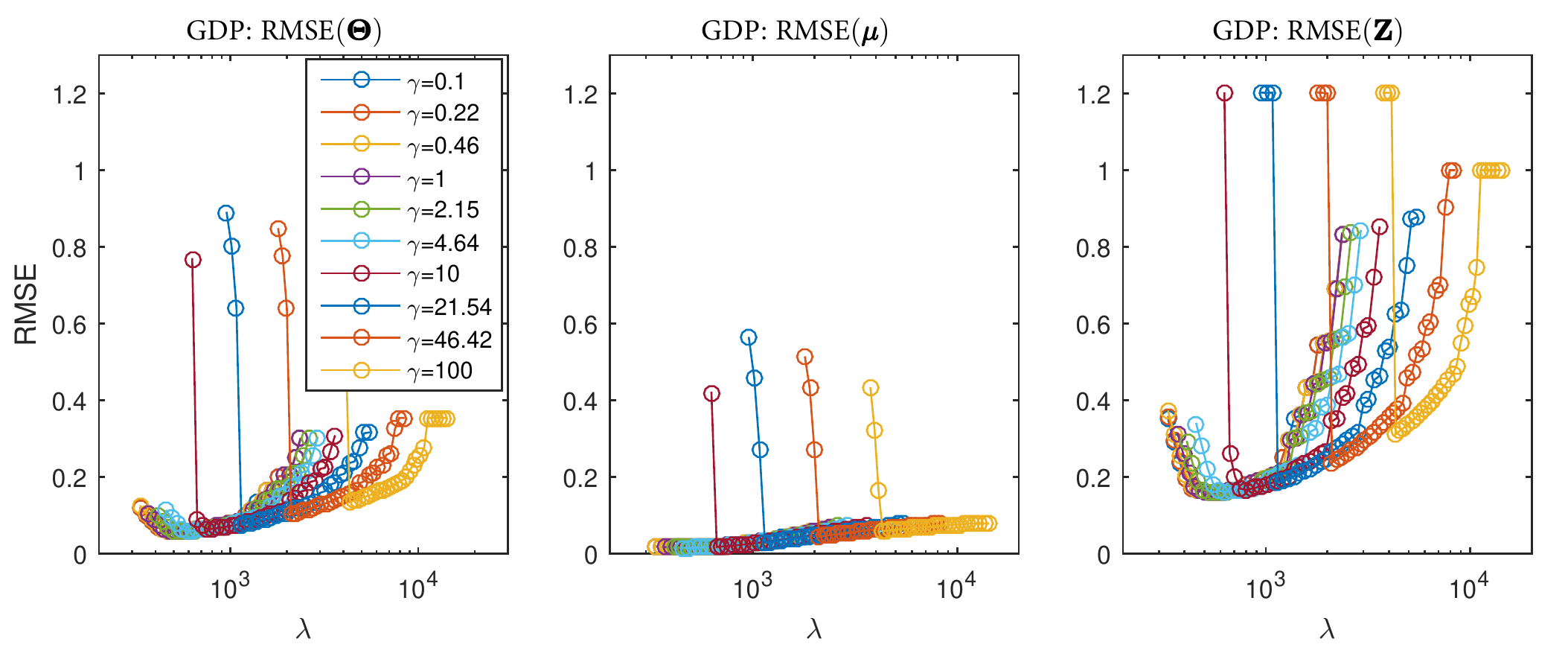}
    \caption*{\textbf{Fig.~S4} Relationship between $\gamma$, $\lambda$ and RMSEs of estimating $\mathbf{\Theta}$, $\bm{\mu}$ and $\mathbf{Z}$ achieved for the GSCA model with GDP penalty. The $x$-axis has a log scale.}
\end{figure}

\begin{figure}[h!]\label{Fig:S5}
    \centering
    \includegraphics[width=\textwidth]{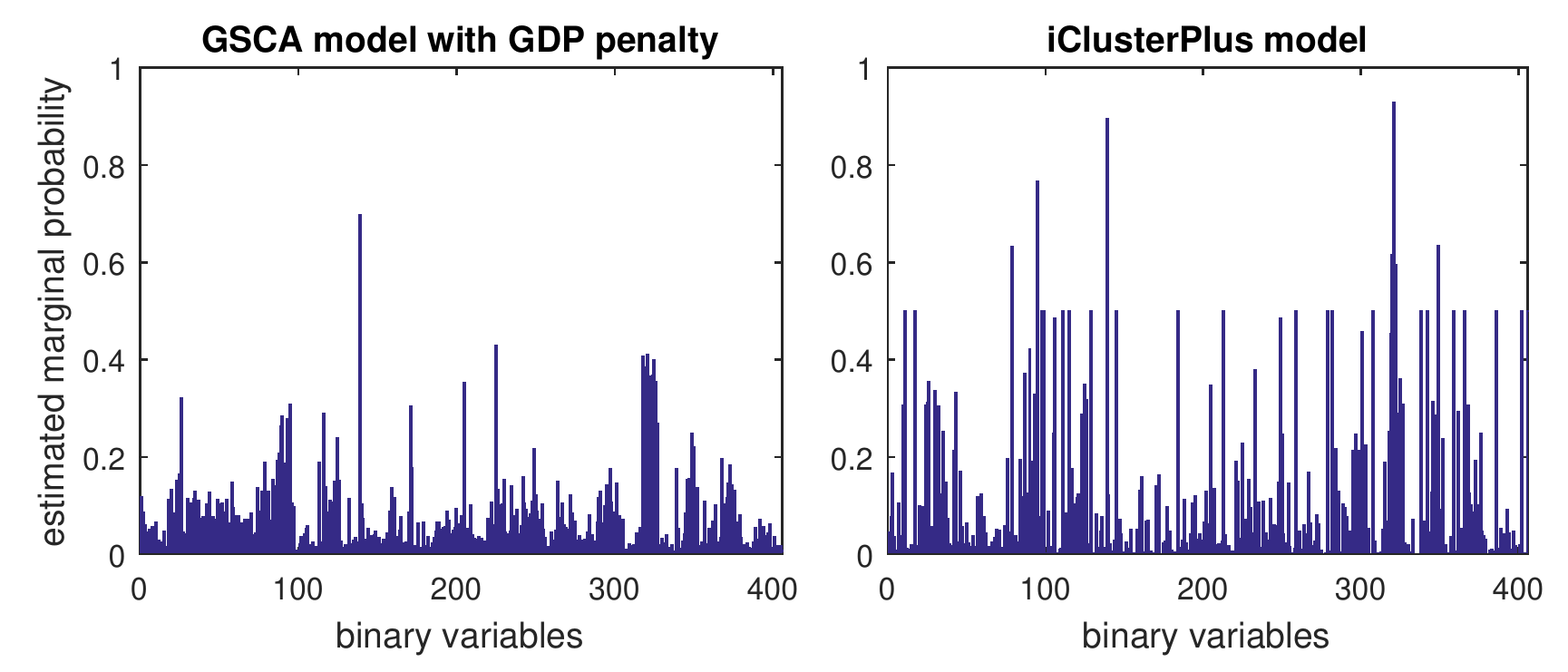}
    \caption*{\textbf{Fig.~S5} Estimated marginal probabilities (the logit transform of the estimated $\hat{\bm{\mu}}_1$) from the GSCA model with GDP penalty (left) and iClusterPlus model (right).}
\end{figure}

\begin{figure}[h!]\label{Fig:S6}
    \centering
    \includegraphics[width=\textwidth]{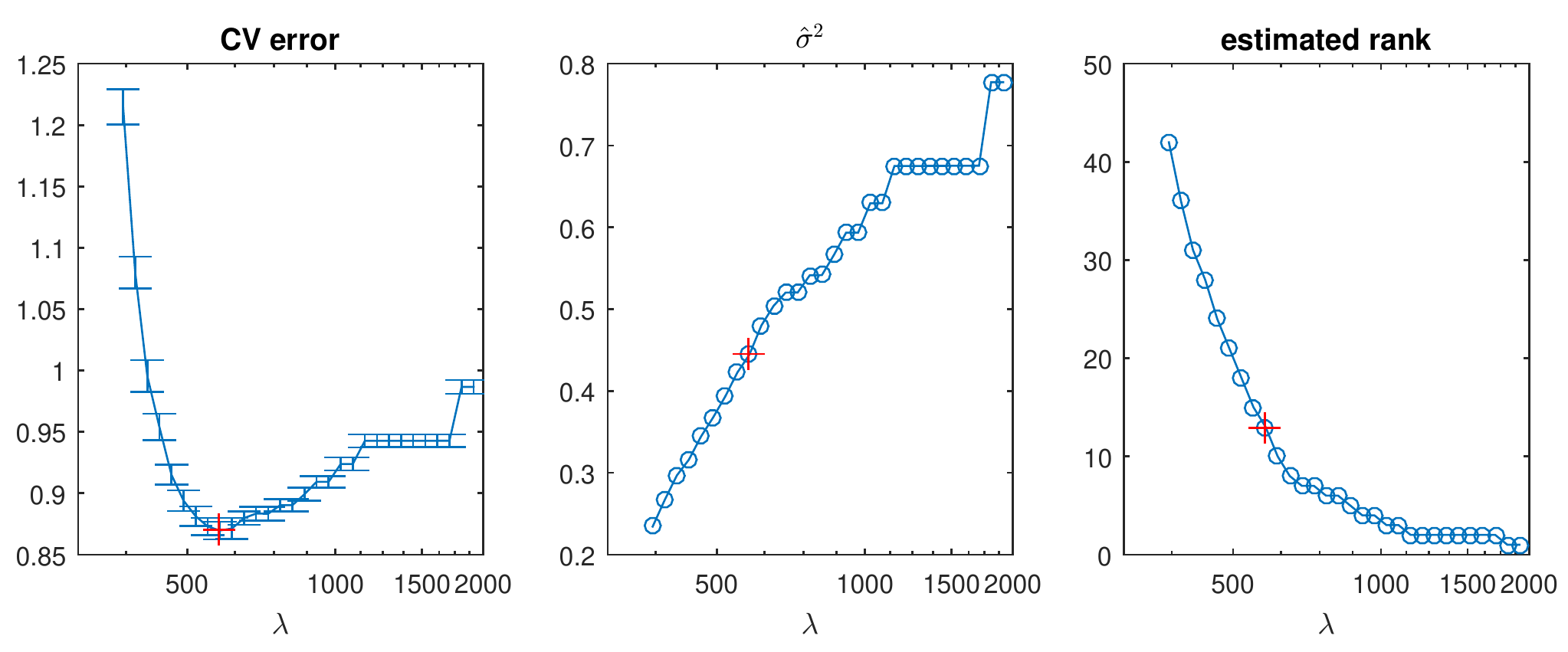}
    \caption*{\textbf{Fig.~S6} Model selection of the GSCA model with GDP penalty on the GDSC data sets: CV error (left); estimated $\hat{\sigma}^2$ (center); $\text{rank}(\hat{\mathbf{Z}})$ (right).}
\end{figure}

\begin{figure}[h!]\label{Fig:S7}
    \centering
    \includegraphics[width= \textwidth]{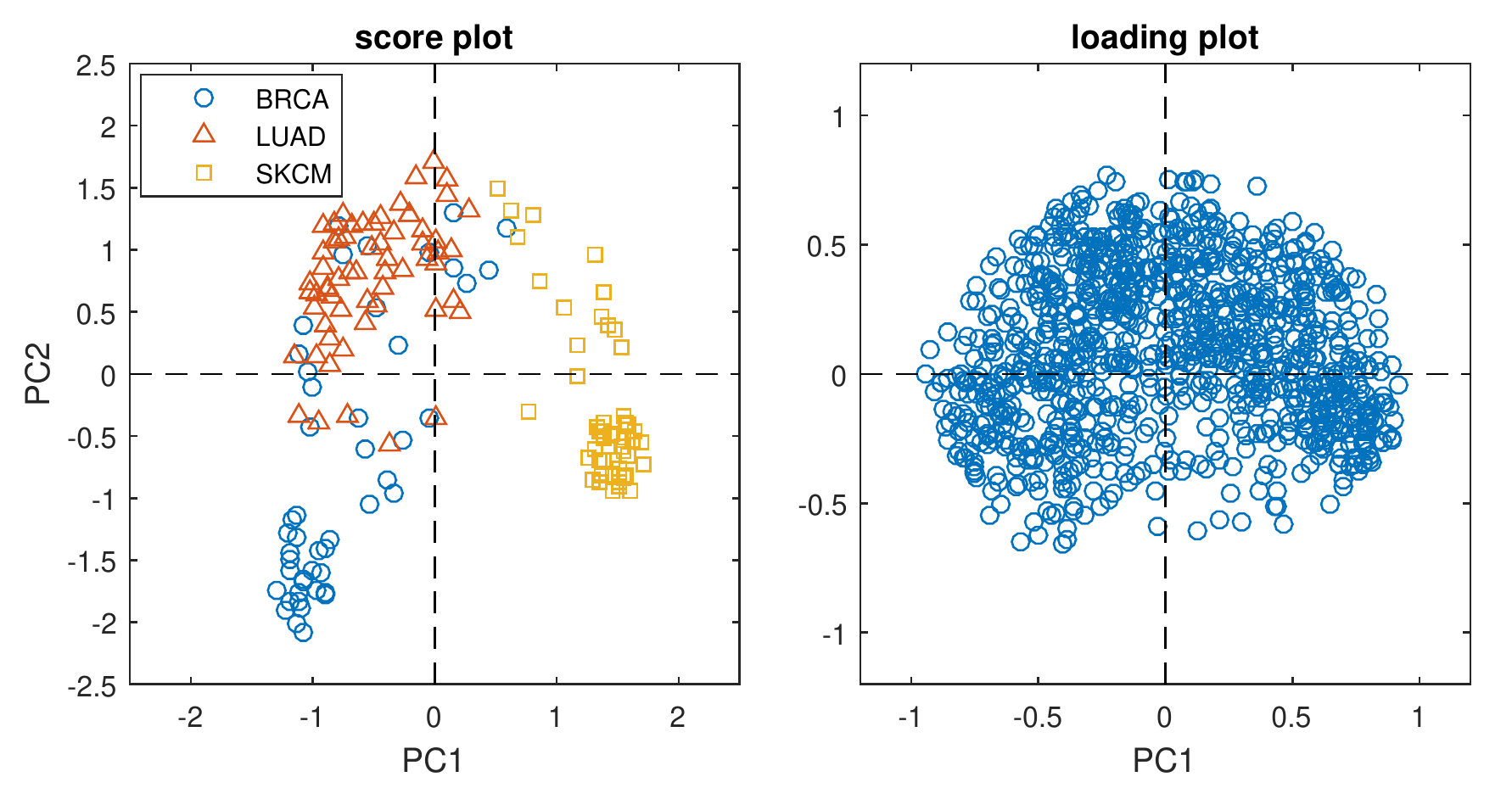}
    \caption*{\textbf{Fig.~S7} Score plot (left) and loading plot (right) derived from a PCA model on the gene expression data $\mathbf{X}_2$. $\mathbf{X}_2$ are centered and scaled in the same as in the GSCA model. SVD algorithm is used to solve the PCA model.}
\end{figure}

\begin{figure}[h!]\label{Fig:S8}
    \centering
    \includegraphics[width= \textwidth]{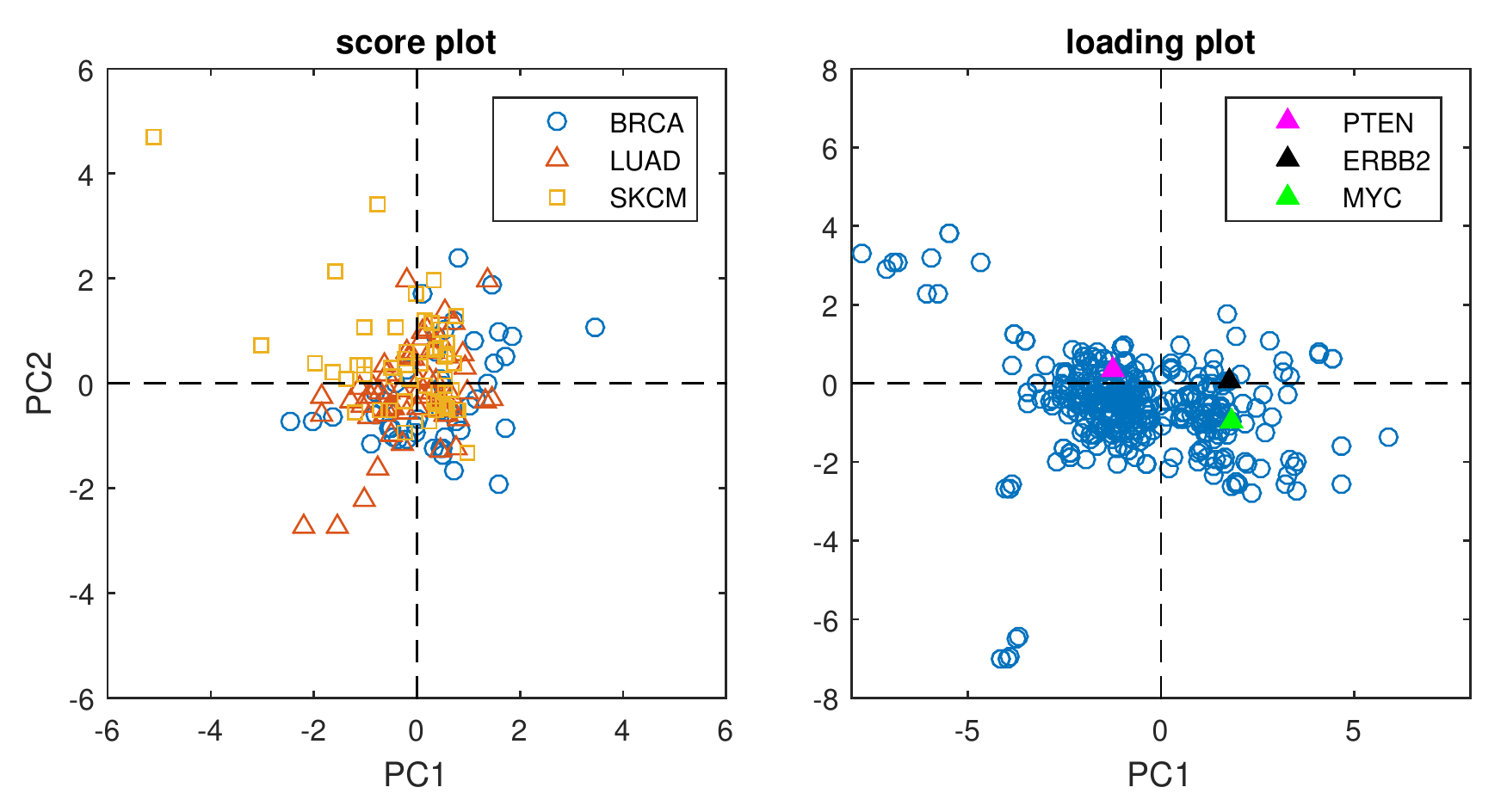}
    \caption*{\textbf{Fig.~S8} Score plot (left) and loading plot (right) are derived from a three components logistic PCA model on the CNA data $\mathbf{X}_1$.}
\end{figure}

\begin{figure}[h!]\label{Fig:S9}
    \centering
    \includegraphics[width=0.5 \textwidth]{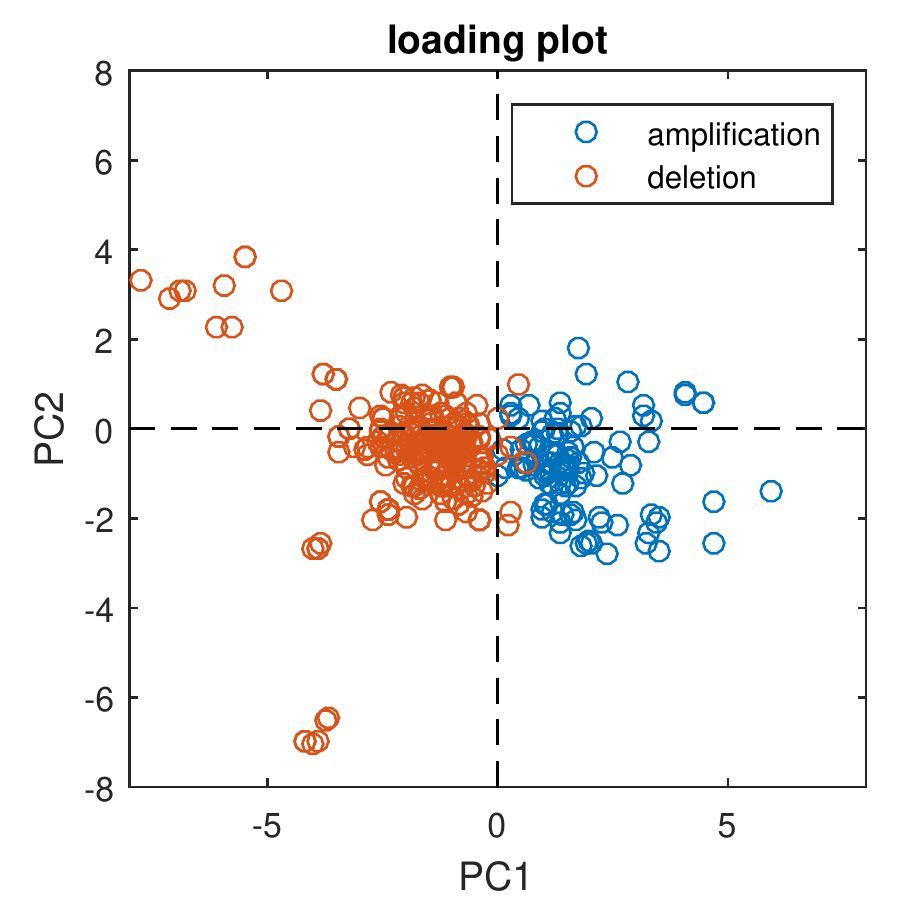}
    \caption*{\textbf{Fig.~S9} Loading plot derived from the three components logistic PCA model on the CNA data $\mathbf{X}_1$. The legend indictates the amplification or deletion of CNA feature.}
\end{figure}

\begin{figure}[h!]\label{Fig:S10}
    \centering
    \includegraphics[width=0.5 \textwidth]{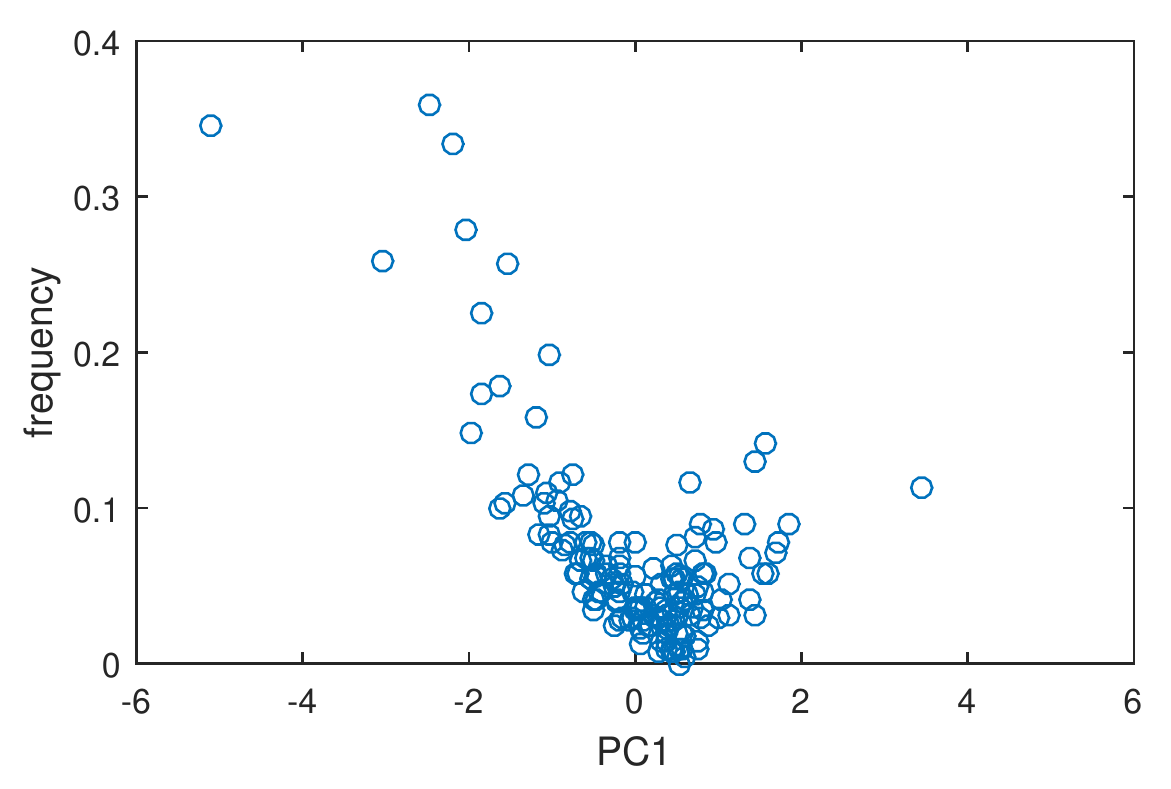}
    \caption*{\textbf{Fig.~S10} The relationship between PC1 scores and the frequency of aberrations of given samples derived from the three components logistic PCA model on the CNA data $\mathbf{X}_1$. }
\end{figure}

\begin{figure}[h!]\label{Fig:S11}
    \centering
    \includegraphics[width= 0.5 \textwidth]{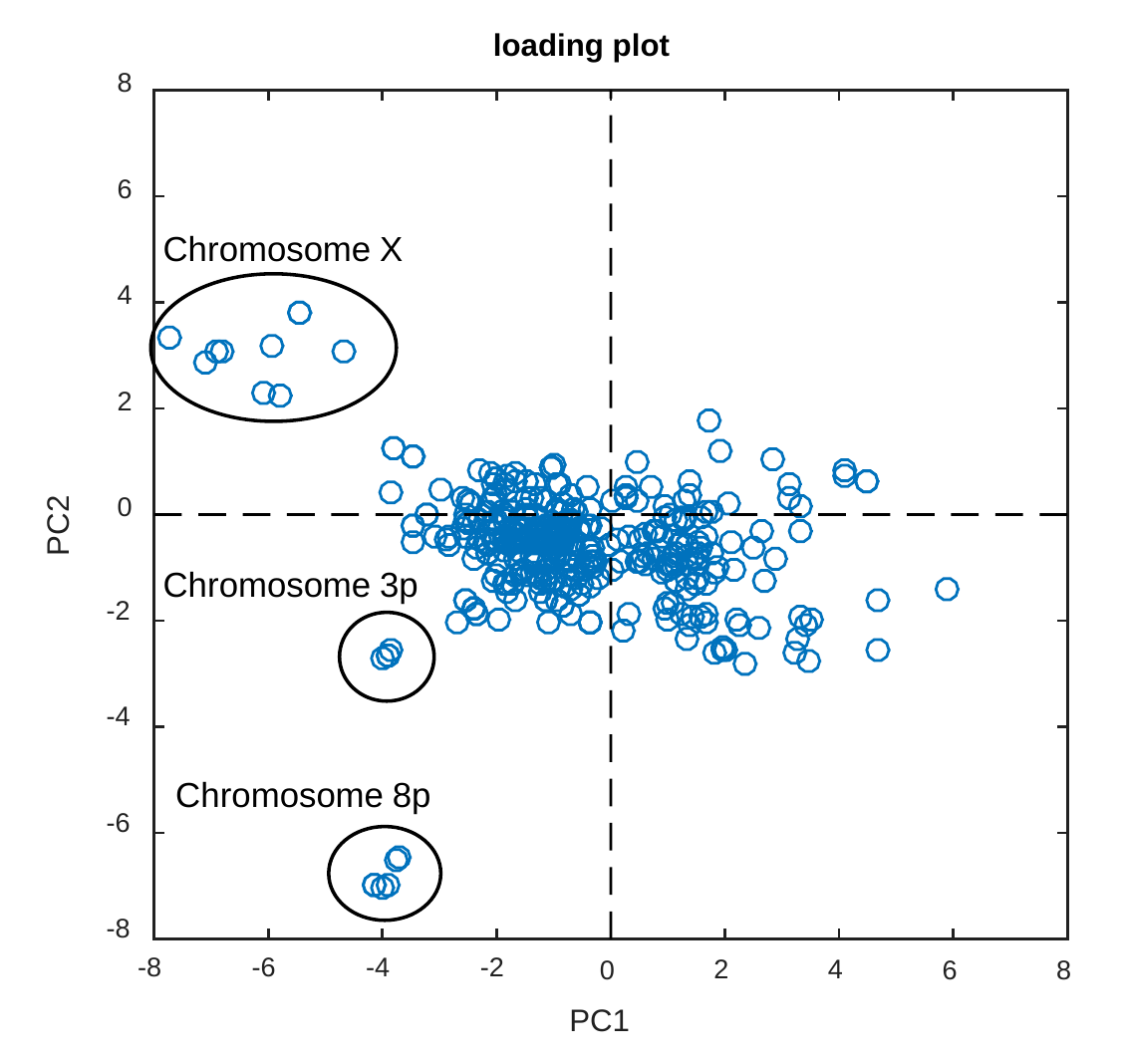}
    \caption*{\textbf{Fig.~S11} Loading plot derived from the three components logistic PCA model on the CNA data $\mathbf{X}_1$. The annotation indicates those features are in the same chromosome region.}
\end{figure}

\begin{figure}[h!]\label{Fig:S12}
    \centering
    \includegraphics[width= \textwidth]{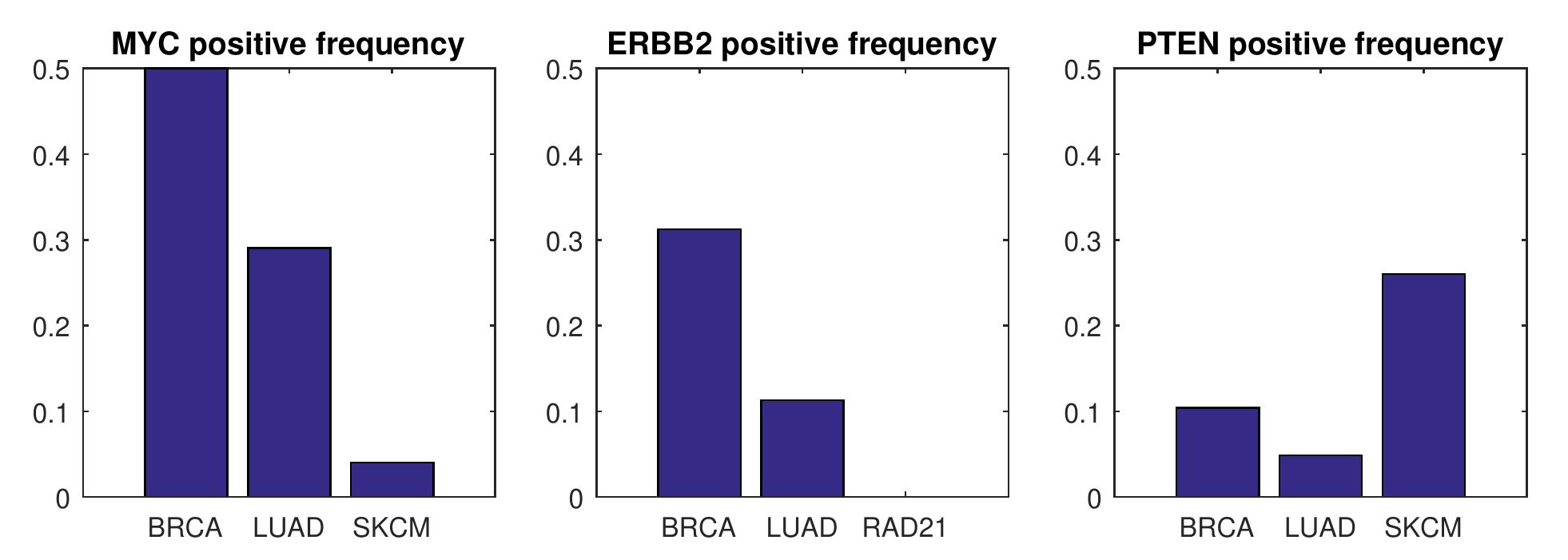}
    \caption*{\textbf{Fig.~S12} Positive frequencies of MYC, ERBB2 and PTEN features in three different cancer types.}
\end{figure}

\end{document}